\documentclass[aps,prd,onecolumn,superscriptaddress,showpacs,showkeys]{revtex4}

\usepackage{slashed}
\usepackage{color}
\usepackage{graphicx}
\usepackage{amsmath}
\usepackage{epsfig}
\def\Comment#1{}
\newcommand{\bean}{\begin{eqnarray*}}
\newcommand{\eean}{\end{eqnarray*}}

\newcommand{\gapproxeq}{\lower
.7ex\hbox{$\;\stackrel{\textstyle >}{\sim}\;$}}
\newcommand{\lapproxeq}{\lower
.7ex\hbox{$\;\stackrel{\textstyle <}{\sim}\;$}}

\newcommand\lsim{\mathrel{\rlap{\lower4pt\hbox{\hskip1pt$\sim$}}
    \raise1pt\hbox{$<$}}}
\newcommand\gsim{\mathrel{\rlap{\lower4pt\hbox{\hskip1pt$\sim$}}
    \raise1pt\hbox{$>$}}}
\newcommand{\ba}{\begin{array}}
\newcommand{\ea}{\end{array}}
\newcommand{\nn}{\nonumber}

\newcommand{\be}{\begin{small}\begin{equation}}
\newcommand{\ee}{\end{equation}\end{small}}
\newcommand{\bear}{\begin{small}\begin{eqnarray}}
\newcommand{\eear}{\end{eqnarray}\end{small}}

\newcommand{\cO}{{\cal O}}

\newcommand{\mA}{\mathcal{A}}

\newcommand{\mR}{\mathcal{R}}

\def\bat{\begin{array}{cc}}

\newcommand{\Frac}[2]{\frac{\displaystyle #1}{\displaystyle #2}}

\newcommand{\bea}{\begin{small}\begin{eqnarray}}
\newcommand{\eea}{\end{eqnarray}\end{small}}

\begin{document}
\begin{flushright}
FTUAM-13-37\\
IFT-UAM/CSIC-13-127
\end{flushright}

% Use the \preprint command to place your local institutional report
% number in the upper righthand corner of the title page in preprint mode.
% Multiple \preprint commands are allowed.
% Use the 'preprintnumbers' class option to override journal defaults
% to display numbers if necessary
%\preprint{}

%Title of paper
\title{{Positivity constraints on the low-energy constants \\
of the chiral  pion-nucleon Lagrangian}   }

% repeat the \author .. \affiliation  etc. as needed
% \email, \thanks, \homepage, \altaffiliation all apply to the current
% author. Explanatory text should go in the []'s, actual e-mail
% address or url should go in the {}'s for \email and \homepage.
% Please use the appropriate macro foreach each type of information

% \affiliation command applies to all authors since the last
% \affiliation command. The \affiliation command should follow the
% other information
% \affiliation can be followed by \email, \homepage, \thanks as well.
\author{Juan~Jos\'e~Sanz-Cillero}
\email[]{juanj.sanz@uam.es}
\affiliation{Departamento de F\'isica Te\'orica and Instituto de F\'isica Te\'orica, IFT-UAM/CSIC
     Universidad Aut\'onoma de Madrid, Cantoblanco, Madrid, Spain}

\author{De-Liang Yao}
\email[]{d.yao@fz-juelich.de}
 \affiliation{Institute for Advanced Simulation, Institut f{\"u}r Kernphysik and
J\"ulich Center for Hadron Physics, Forschungszentrum J{\"u}lich, D-52425 J{\"u}lich, Germany}

\author{Han-Qing~Zheng}
\email[]{zhenghq@pku.edu.cn}
\affiliation{Department of Physics and State Key Laboratory of
Nuclear Physics and Technology, Peking University, Beijing 100871,
People's Republic of China}

%Collaboration name if desired (requires use of superscriptaddress
%option in \documentclass). \noaffiliation is required (may also be
%used with the \author command).
%\collaboration can be followed by \email, \homepage, \thanks as well.
%\collaboration{}
%\noaffiliation

%%%\date{\today}

\begin{abstract}
Positivity constraints on the pion-nucleon scattering amplitude are derived in this article
with the help of general S-matrix arguments, such as analyticity, crossing symmetry and unitarity,
in the upper part of Mandelstam triangle, ${\cal R}$. Scanning inside the region ${\cal R}$,
the most stringent bounds on the chiral low energy constants of the pion-nucleon Lagrangian
are determined. When just considering the central values of the fit results from
covariant baryon chiral perturbation theory using extended-on-mass-shell scheme,
it is found that these bounds are well respected numerically both at the $O(p^3)$ and $O(p^4)$ level.
Nevertheless, when taking the errors into account, only the $O(p^4)$ bounds are obeyed
{
in the full error interval, while the bounds on
$O(p^3)$  fits are slightly violated.
If one disregards loop contributions, the bounds always fail in certain regions of ${\cal R}$. Thus,
at a given chiral order these terms are not numerically negligible and one needs to consider
all possible contributions, {\it i.e.},  both tree-level  and loop diagrams.
We have provided the constraints for special points in $\mR$  where the bounds are nearly optimal
in terms of just a few chiral couplings, which can be easily implemented
and employed to constrain future analyses.
Some issues about calculations with an explicit $\Delta$ resonance are also discussed.
}
\end{abstract}

% insert suggested PACS numbers in braces on next line
\pacs{12.39.Fe, 11.55.Fv}
% insert suggested keywords - APS authors don't need to do this
\keywords{Pion-nucleon scattering, Positivity constraints}

%\maketitle must follow title, authors, abstract, \pacs, and \keywords
\maketitle

\section{Introduction}

Chiral perturbation theory ($\chi$PT )~\cite{ChPT} plays an important role in studying low energy hadron physics, such as the pion-nucleon interaction. Many efforts have been made to study pion-nucleon physics within baryon chiral perturbation theory (B$\chi$PT)~\cite{gasser} using different approaches,  e.g., heavy baryon (HB) $\chi$PT~\cite{HB}, infrared regularization (IR)~\cite{IR},
extended on mass shell (EOMS)~\cite{EOMS}, etc. The scattering amplitudes are then expressed
in terms of the low energy constants (LECs). As   {it} is well known,
when stepping up to higher {and higher} orders,
there always appears a {rapidly}   growing number {of LECs,}
which are free parameters,   not fixed by chiral symmetry. Nevertheless, general S-matrix arguments
such as analyticity, crossing and unitarity
can be used to constrain the pion-nucleon interaction and its chiral effective theory description. It is therefore possible
to obtain certain {model-independent}
constraints on the LECs.

Along this {line}, many works have been devoted to { the study of }
positivity constraints
on $\pi\pi$ scattering amplitudes
(e.g., see Refs.~\cite{Ananthanarayan,Dita,Manohar,Mateu,Guo}).
{
The pion-nucleon scattering was also studied in Ref.~\cite{Luo},
in terms of the pion energy $E_\pi$ in the center-of-mass rest-frame
%%%
%%%the variable $E_\pi=(s-m_N^2+M_\pi^2)/(2\sqrt{s})$
%%%
and
positivity constraints were extracted for
the second derivative of the
%%%
%%$\frac{d^2D(s,t)}{ds^2}^{\rm non-pole}$ for
$\pi^\pm p\to \pi^\pm p$ scattering amplitude with respect to $E_\pi$.
%%%
%%%where the nucleon pole was explicitly subtracted.
%%%
However,  only the $\pi^+ p$ forward scattering ($t=0$)  was  analyzed in detail  and
no extra information was extracted from  the $\pi^- p$ channel.
Likewise, the positivity of its second derivative was only analyzed at two particular points,
$E_\pi=\pm M_\pi/\sqrt{2}$~\cite{Luo}.
The central values from  HB-$\chi$PT~\cite{HBp3} were employed to check the obtained bounds.
}

In this paper, the analysis is extended beyond the forward case $t=0$
to the full upper part of the Mandelstam triangle ${\cal R}$ (with $t>0$).
In Sec.~\ref{sec.piN-scat} we introduce the general properties of pion-nucleon scattering.
{A particular combination $D_\alpha$  of the pion-nucleon scattering functions $A(s,t)$ and $B(s,t)$
%% $D_\alpha=\alpha D+(1-\alpha)\nu B$
is written down {in terms of a}
positive definite spectral function in Sec.~\ref{sec.specfun}. {It is then used }
to extract
the positivity constraints {for both  $\pi^\pm p\to \pi^\pm p$ scatterings}
in Sec.~\ref{sec.theory}.
Hence, compared to Ref.~\cite{Luo},  extra   information coming from the $B(s,t)$ function
and the $\pi^- p\to\pi^- p$ scattering is taken into consideration
{in the present work.   }
Rather than taking two particular points to get two  bounds,
we scan the full region ${\cal R}$, extracting the most stringent bounds on the LECs.
These are then tested in Sec.~\ref{sec.pheno} by means of the recent results from
relativistic B$\chi$PT using EOMS scheme~\cite{oller,yao}.
%%%
%%%It is better to adopting the EOMS-B$\chi$PT results
%%%
This scheme is more convenient for our analysis than the HB$\chi$PT ones,
as EOMS-B$\chi$PT possesses the correct analytic behaviour  in the  Mandelstam triangle.
The uncertainties due to the LEC errors and the impact of the $\Delta$ resonance are also analyzed
in Sec.~\ref{sec.pheno}.
The conclusions are summarized in Sec.~\ref{sec.conclusions}
and some technical details about the positivity of the right-hand cut spectral function are relegated
to the Appendix.
}

\section{Aspects of elastic pion-nucleon scattering}
\label{sec.piN-scat}

%\subsection{Effective Lagrangian and kinematics}
The effective Lagrangian describing {the}  low-energy pion-nucleon scattering at $O(p^4)$ level takes the following form:
\bea
{\cal L}_{\pi N}&=&\bar{\Psi}\left\{i\,\slashed{D}-m+\frac{g}{2}\,\slashed{u}
\gamma_5+\sum_{i=1}^7c_i\,{\cal O}^{(2)}_i+\sum_{j=1}^{23}d_j\,{\cal O}^{(3)}_j+\sum_{k=1}^{118}e_k\,{\cal O}^{(4)}_k\right\}\Psi+\cdots
\eea
where ${\cal O}_i^{(m)}$s ($m=2,3,4$) are the  operators
of $O(p^m)$. {Their}   explicit expressions can be found in Ref.~\cite{eff} and the references therein.
Here $m$ and $g$ denote the nucleon mass and the axial charge in the chiral limit.
The coefficients $c_i, d_j, e_k$ are {LECs}, given in  units of GeV$^{-1}$, GeV$^{-2}$
and GeV$^{-3}$, respectively.

In the isospin limit, the scattering amplitude for the process of $\pi^a(q)+N(p)\to\pi^{a^\prime}(q^\prime)+N(p^\prime)$ with isospin indices $a$ and $a^\prime$ is described by $A^\pm(s,t)$, $B^\pm(s,t)$ and $D^\pm(s,t)$ according to~\cite{gasser,pinbecher}
\bea \label{ABform}
T_{\pi N}^{a^\prime a}(s,t)&=& \chi_{N'}^\dagger \,\bigg\{  \,
\frac{1}{2}\{\tau_{a^\prime},\tau_a\}T^+(s,t)+\frac{1}{2}[\tau_{a^\prime},\tau_a]T^-(s,t)
\, \bigg\}\, \chi_N  \ , \\
T^{\pm}(s,t)&=&\bar{u}(p^\prime)\left[A^\pm(s,t)+\frac{\slashed{q}^\prime+\slashed{q}}{2}B^\pm(s,t)\right]u(p)=\bar{u}(p^\prime)\left[D^\pm(s,t)+\frac{[\slashed{q}^\prime,\slashed{q}]}{4m_N}B^\pm(s,t)\right]u(p)\ ,\\
D^\pm(s,t)&=&A^\pm(s,t)+\nu B^\pm(s,t)\ ,
\eea
here $\tau_{a^\prime}$, $\tau_{a}$ are Pauli matrices,  $\nu=(s-u)/{4m_N}$
{ and $\chi_N$ ($\chi_{N'}$) is the isospinor for the incoming (outgoing) nucleon.
}The Mandelstam variables s, t and u fulfill $s+t+u=2m_N^2+2M_\pi^2$ with $m_N$ and $M_\pi$,
{being}     the physical {  nucleon and pion masses, respectively. }
The functions $X^\pm$ with $X=\{A,B,D\}$ are the so-called isospin-even (for `+') and -odd (for `-')
amplitudes, and they are related to the isospin amplitudes with definite isospin $I$ ($1/2$ or $3/2$) via
\bea
X^{1/2}=X^++2X^-\ ,\qquad X^{3/2}=X^+-X^-\ .
\eea
It is also convenient for later use to write down the relations among the $\pi^\pm p\to\pi^\pm p$
scattering amplitudes , isospin even/odd amplitudes and isospin {amplitudes:}
\bea\label{relationamong}
X^{\pi^+p}\, =\, X^{3/2}\, =\, X^+ \,  - \, X^-\ ,\qquad \qquad
X^{\pi^-p}\, = \, \frac{2}{3}X^{1/2}\, + \, \frac{1}{3}X^{3/2}\, = \, X^+ \, +  \, X^-\ .
\eea

\begin{figure}[!t]
\centering
\centering
\includegraphics[clip,width=0.4\textwidth]{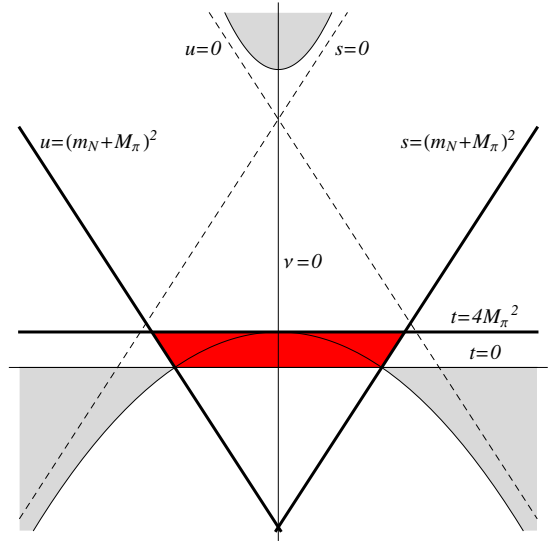}
\caption{\label{mandel}
{\small Mandelstam  plane $(\nu,t)$.
The Mandelstam triangle is the region contoured by the $s=(m_N+M_\pi)^2$, $u=(m_N+M_\pi)^2$
and $t=4 M_\pi^2$ lines. Our region of study $\mR$ is the trapezium  formed
by the three previous lines and $t=0$, which is marked in red.
}
}
\end{figure}

The physical region for the pion-nucleon reaction corresponds to the kinematical region where the Kibble
function~\cite{Kibble} $\Phi=t\left[su-(m_N^2-M_\pi^2)^2\right]$ is non-negative.
In Fig.~\ref{mandel}, the physical regions are depicted by light gray.{
The triangle
%%%(blue)
in the center is  given  by $s,\, u \leq (m_N+M_\pi)^2$ and $t<4 M_\pi^2$.
{It  is}   the so-called
Mandelstam triangle. The upper part of the Mandelstam triangle
bounded {by $t\geq 0$} corresponds to the region ${\cal R}$ (marked in red in Fig.~\ref{mandel})
where the positivity  conditions are considered.
In terms of the $(\nu,t)$ variables the Mandelstam diagram is given by $t\leq 4 M_\pi^2$
and $|\nu |\leq \nu_{th}(t) = M_\pi +t/(4 m_N)$. In order to obtain the region  $\mR$
one should add the restriction $t\geq 0$.
}

\section{Partial wave decomposition and positive definite spectral function\label{sec.specfun}}

It is well known that the full isospin amplitude can be written in terms of the partial-wave (PW) amplitudes as~\cite{roy-steiner}
\bea
\vec{{\cal A}}^I(s,t)\bigg|_{t=t(s,z_s)}=\sum_{\ell=0}^{\infty}S^\ell(s,z_s)\vec{{\cal F}}_\ell^I(s)\ ,
\eea
with
\bea
\vec{{\cal A}}^I\equiv\left(
                        \begin{array}{c}
                          A^I \\
                          B^I \\
                        \end{array}
                      \right)
\ ,\quad
\vec{F}^I_\ell\equiv\left(
                      \begin{array}{c}
                        f_{\ell+}^I \\
                        f_{(\ell+1)-}^I \\
                      \end{array}
                    \right)
\ ,\nonumber
\eea and
\bea\label{kernelS}
S^\ell(s,z_s)=4\pi\left(
                    \begin{array}{cc}
                      \left[\frac{W+m_N}{E+m_N}P_{\ell+1}^\prime(z_s)+\frac{W-m_N}{E-m_N}P_{\ell}^\prime(z_s)\right] & -\left[\frac{W+m_N}{E+m_N}P_{\ell}^\prime(z_s)+\frac{W-m_N}{E-m_N}P_{\ell+1}^\prime(z_s)\right] \\
                      \left[\frac{1}{E+m_N}P_{\ell+1}^\prime(z_s)-\frac{1}{E-m_N}P_{\ell}^\prime(z_s)\right] & -\left[\frac{1}{E+m_N}P_{\ell}^\prime(z_s)-\frac{1}{E-m_N}P_{\ell+1}^\prime(z_s)\right] \\
                    \end{array}
                  \right)\ , \quad W=\sqrt{s}.
\eea
Here $P_{\ell}(z_s)$ are the conventional Legendre polynomials and $z_s=1+\frac{2s\,t}{\lambda(s,m_N^2,M_\pi^2)}$ with $\lambda(s,m_N^2,M_\pi^2)=[s-(m_N+M_\pi)^2][s-(m_N-M_\pi)^2]$,
{is}
the K$\ddot{a}$ll$\acute{e}$n function.
These set of kernel matrices $S^\ell(s,z_s)$ are always analytical functions, real for real values of the
Mandelstam variables $(s,t,u)$. Thus, in the case $s\geq s_{th}$ the whole analytic discontinuity is
due to the partial waves $f^I_k(s)$:
\bea\label{imageexp}
{\rm Im}\vec{\cal A}^I(s+i\epsilon,t) =
\sum_{\ell=0}^\infty S^\ell(s,z_s(s,t)) \ {\rm Im}\vec{\cal F}^I(s+i\epsilon)\ .
\eea
Since a fixed--$t$ dispersion relation for the analysis of the subthreshold amplitude will be used in Sec.~\ref{secdisper},
our interest is focused on obtaining a positive definite spectral function in the physical region
$s\geq s_{th}$. On the right-hand side of Eq.~(\ref{imageexp}), the imaginary part of each PW is positive due to unitarity, i.e., ${\rm Im}f_k^I(s)\geq0$ for $s\geq s_{\rm th}$, but the kernel matrices always contain negative elements. Therefore, it is proper to construct a combination of $A^I$ and $B^I$ in the form
\bea
D_\alpha^I(s,t)\equiv \alpha A^I(s,t)+\nu B^I(s,t)=\alpha D^I(s,t)+(1-\alpha)\nu B^I(s,t)
\eea
such that its imaginary part satisfies
\bea\label{specfun}
{\rm Im}D_\alpha^I(s,t)\geq0\ .
\eea
In order to guarantee Eq.~(\ref{specfun}), it is
{proven} in great detail in App.~\ref{secspecfun} that the validity region for the combination factor $\alpha$ should be $\alpha_{\rm min}(t)\leq\alpha\leq\alpha_{\rm max}(t)$ with
\bea
\alpha_{\rm min}(t)\, =\,  \Frac{ \left(1 + \frac{t}{4 m_N M_\pi} \right)\,
\left(1-\frac{t}{4 m_N^2} \right)
}{  \left( 1+ \frac{t}{2 m_N M_\pi} +\frac{t}{4 m_N^2}\right)} \, =\, 1 \, -\, \Frac{t}{4 m_N M_\pi}
\, +\, \cO\bigg(\Frac{p^2}{m_N^2}\bigg) \, ,
\qquad \qquad
\alpha_{\rm max}(t) \,= \, 1+\frac{t}{4 m_N M_\pi} \, ,
\eea
where $M_\pi=\cO(p)$ and $t=\cO(p^2)$~\cite{oller,yao}.

 It is worth noting that here the Mandelstam variable $t$ must be greater than zero, i.e., $t\geq0$,  due to the application of Eq.~(\ref{legendre}) and the fact of $P_k^\prime(z_s)\geq0$ for $z_s\geq1$ in App.~\ref{secspecfun}. This is the reason why our analysis of the positivity constraints is restricted to the upper part of the Mandelstam triangle ${\cal R}$ (see Fig~\ref{mandel}).

So far, the $s$-channel positive definite spectral function above threshold
is clear. The corresponding $u$-channel one is easily obtainable
by crossing symmetry{:}
\bea\label{crossing}
D_\alpha^I(u,t)=C_{LR}^{II^\prime}D_\alpha^{I^\prime}(s,t)\ ,\quad\text{(or, equivalently, $D_\alpha^I(-\nu,t)=C_{LR}^{II^\prime}D_\alpha^{I^\prime}(\nu,t)$)}
\eea
with the crossing matrix being
\bea
C_{LR}=\frac{1}{3}\left(
         \begin{array}{cc}
           -1 & 4 \\
           2 & 1 \\
         \end{array}
       \right)\ ,\quad C_{LR}^{II^\prime}C_{LR}^{I^\prime J}=\delta_{IJ}\ .
\eea
where the first (second)  row and column of $C_{LR}$ correspond to isospin $\frac{1}{2}$
(isospin $\frac{3}{2}$). $C_{LR}$ can be also sometimes denoted in the bibliography as $C_u$.

\section{\label{secdisper}Theoretical constraints indicated by the dispersion relation}
\label{sec.theory}

For $0\leq t \leq 4 M_\pi^2$ it is possible to write down a fixed--$t$ dispersion relation for
{the $X(\nu,t)$}  in terms of the $\nu$ variable (or $s$, if desired).
If    {   $\nu D^I_\alpha(\nu,t) $   }
vanished for $|\nu|\to \infty$,
the amplitude $D_\alpha^I(\nu,t)$ could be represented then by the unsubtracted
dispersive integral,
\bea
\label{dispersion}
D_\alpha^I(\nu,t)&=&\frac{Z_{N,R}^I(t)}{\nu_B-\nu}+\frac{Z_{N,L}^I(t)}{\nu_B+\nu}+\frac{1}{\pi}\int_{\nu_{th}}^\infty d\nu^\prime\left[
\frac{{\rm Im}D_\alpha^{I}(\nu^\prime+i\epsilon,t)}{\nu\prime-\nu}+\frac{{\rm Im}D_\alpha^{I}(-\nu^\prime-i\epsilon,t)}{\nu\prime+\nu}\right]\ ,
\eea
where
{$\nu_B(t) =\nu|_{s=m_N}= (t-2 M_\pi^2)/(4 m_N)$   and  }
 $Z_{N,R}^I(t)$ and $Z_{N,L}^I(t)$ are the residues
 of the s- and u-channel nucleon poles,
respectively. {The} first term {within}
the integral {comes}  from the discontinuity across the right-hand cut,
and the second {one}  from the {discontinuity}
across the left-hand cut. Since the left-hand {cut}  spectral
function  {Im$D_\alpha^{I}(-\nu^\prime-i\epsilon,t)$}
with isospin $I$ and the right-hand spectral
function   {Im$D_\alpha^{I^\prime}(\nu^\prime+i\epsilon,t)$}
with isospin $I^\prime$ are related by
the crossing relation {in}  Eq.~(\ref{crossing}),
the dispersion relation~(\ref{dispersion}) can be rewritten as
%%%
%%%\bea\label{dispersion2}
%%%D_\alpha^I(\nu,t)&=&\frac{Z_{N,R}^I(t)}{\nu_B-\nu}+\frac{Z_{N,L}^I(t)}{\nu_B+\nu}
%%%+\frac{1}{\pi}\int_{\nu_{th}}^\infty d\nu^\prime\left[\frac{\delta^{II^\prime}}{\nu\prime-\nu}
%%%+\frac{C_{LR}^{II^\prime}}{\nu\prime+\nu}\right]{\rm Im}D_\alpha^{I\prime}(\nu^\prime+i\epsilon,t)\ .
%%%\eea
%
{
\bea\label{dispersion2}
\tilde{D}_\alpha^I(\nu,t)&=&
\frac{1}{\pi}\int_{\nu_{th}}^\infty d\nu^\prime\left[\frac{\delta^{II^\prime}}{\nu\prime-\nu}
+\frac{C_{LR}^{II^\prime}}{\nu\prime+\nu}\right]{\rm Im}D_\alpha^{I\prime}(\nu^\prime+i\epsilon,t)\ .
\eea
with the nucleon pole subtracted amplitude,
\bea
\tilde{D}_\alpha^I(\nu,t)
\,\,\, \equiv \,\,\,
D_\alpha^I(\nu,t)-\left[\frac{Z_{N,R}^I(t)}{\nu_B-\nu}+\frac{Z_{N,L}^I(t)}{\nu_B+\nu}\right]\, .
\eea
}

{
In the physical case, however, $\nu D_\alpha(\nu,t)$ does not vanish at high energies
and the unsubtracted dispersive integral in Eq.~(\ref{dispersion2})
does not converge.  Nonetheless,   {this} can be easily cured by considering a number
$n\geq 2$ of subtractions. An equivalent alternative is to
take the $n$-th derivative with respect to $\nu$ on both sides
{of Eq.~(\ref{dispersion2})~\cite{roy-steiner,disp-relations}:            }
\bea\label{dispersion3}
\frac{d^n}{d\nu^n}\tilde{D}_\alpha^I(\nu,t)=\frac{n!}{\pi}\int_{\nu_{th}}^\infty
d\nu^\prime\left[\frac{\delta^{II^\prime}}{(\nu\prime-\nu)^{n+1}}
+(-1)^n\frac{C_{LR}^{II^\prime}}{(\nu\prime+\nu)^{n+1}}\right]
{\rm Im}D_\alpha^{I^\prime}(\nu^\prime+i\epsilon,t)\ ,
\eea
which is now convergent   {for $n\geq 2$.   }
An analogous  expression  is given  for the $\pi\pi$--scattering amplitude
in Ref.~\cite{Manohar}.
%%
%%In Eq.~(\ref{dispersion3}),
On the right-hand cut ($\nu >\nu_{th}$),
the spectral functions
Im$D_\alpha^{I^\prime}(\nu^\prime+i\epsilon,t)$ are  positive for $\alpha$ in the range
\bear
\alpha_{\rm min}(t) \,\,\, \leq \,\,\, \alpha\,\,\, \leq \,\,\, \alpha_{\rm max}(t)\, .
\eear

Both denominators within the bracket in  Eq.~(\ref{dispersion3})
happen to be   positive for $\nu^\prime\geq\nu_{th}$ when $|\nu|\leq \nu_{th}$.
}
If $n$ is an even number, the relative sign is also positive.
However, { the factor $C_{LR}^{II^\prime}$ is negative when $I=I'=1/2$.   }
The aim, therefore, is to construct combinations of isospin amplitudes in the form
\bea
\sum_I\beta_I\tilde{D}_\alpha^I=\beta_{1/2}\tilde{D}_\alpha^{1/2}+\beta_{3/2}\tilde{D}_\alpha^{3/2}\ ,
\eea
such that both their right- and left-cut contributions are positive-definite.
The inspection of Eq.~(\ref{dispersion3}) implies the constraints
\bea
\sum_I\beta_I\delta^{II^\prime}\geq0\ ,\qquad \sum_I\beta_IC_{LR}^{II^\prime}\geq0\ ,
\eea
which lead to
\bea
2\beta_{3/2}\geq\beta_{1/2}\geq0\ .
\eea
As pointed out by Ref.~\cite{Luo}, it is only necessary to investigate two cases: $\tilde{D}_\alpha^{3/2}$ and $(2\tilde{D}_\alpha^{1/2}+\tilde{D}_\alpha^{3/2})/3$. In view of Eq.~(\ref{relationamong}), they correspond to the physical processes $\pi^+p\to\pi^+p$ and $\pi^-p\to\pi^-p$ respectively.
Hence,
%%%
%%%in combination of Eqs.~(\ref{relationamong}) and~(\ref{dispersion3}),
%%%
two positivity constraints on the pion-nucleon scattering amplitudes are obtained:
\bea
\frac{{\rm d}^n\tilde{D}_\alpha^{\pi^\pm p}(\nu,t)}{{\rm d}\nu^n}
=\frac{{\rm d}^n}{{\rm d}\nu^n}\left[\tilde{D}_\alpha^+(\nu,t)
\mp \tilde{D}_\alpha^-(\nu,t)\right]\geq0\quad\text{(for even $n$)}\ .
\eea
The inequalities above are equivalent to
\bea\label{consonampl}
\frac{{\rm d}^n}{{\rm d}\nu^n}\tilde{D}_\alpha^+(\nu,t)
\,\, \, - \,\, \,   \left|\frac{{\rm d}^n}{{\rm d}\nu^n}\tilde{D}_\alpha^-(\nu,t)\right|
\,\,\, \geq \,\,\, 0\, ,   \qquad(\nu,t)
\in {\cal R}\quad\text{(for even $n$)}\ ,
\eea
and  $\alpha_{\rm min}(t)\leq\alpha\leq\alpha_{\rm max}(t)$. From now one we will focus just on the
$n=2$ case and for later convenience we will define the quantity
\bea
f(\alpha,\nu,t)\,\,\,=\,\,\, F_\pi^2 \, \frac{{\rm d}^2}{{\rm d}\nu^2}\tilde{D}_\alpha^+(\nu,t)
\,\, \, - \,\, \,  F_\pi^2 \,  \left|\frac{{\rm d}^2}{{\rm d}\nu^2}\tilde{D}_\alpha^-(\nu,t)\right|
\, ,
\label{eq.chiral-ineq}
\eea
which must be positive for $(\nu,t)\in {\cal R}$ and
$\alpha_{\rm min}(t)\leq\alpha\leq\alpha_{\rm max}(t)$.

{
Notice that $t=0$ corresponds to the forward scattering case {where, then, }
$\alpha=\alpha_{\rm min}(0)=\alpha_{\rm max}(0)=1$ and $\tilde{D}_1(\nu,t)=\tilde{D}(\nu,t)$.
}
This case was considered in Ref.~\cite{Luo} within the HB-$\chi$PT framework.
In the present work, the analysis has been extended to the  much   wider region ${\cal R}$
in order to obtain more stringent positivity constraints. Moreover,
the recent covariant EOMS-B$\chi$PT {results~\cite{oller,yao}  }
are adopted to test the resultant bounds  on the LECs.

\section{Numerical analysis of the positivity constraints within EOMS-B$\chi$PT}
\label{sec.pheno}

The positivity conditions on {the}
pion-nucleon scattering amplitude, shown in Eq.~(\ref{consonampl}),
can be transformed into bounds on the LECs. By considering the fit results from  B$\chi$PT
one can  test  whether the bounds are respected or not at a given chiral order.
However, as mentioned in Ref.~\cite{Luo},
the scattering amplitudes within HB$\chi$PT manifest an incorrect analytic behavior
inside the Mandelstam triangle,{
e.g., a modification of the nucleon pole structure, which causes  problems with the convergence
of chiral expansion. Hence, it is convenient  to adopt the recent
{relativistic results from the EOMS-B$\chi$PT framework,  }
employed in Refs.~\cite{oller} (up to $O(p^3)$) and~\cite{yao} (up to $O(p^4)$).
}
In what follows, the case $n=2$ {given by Eq.~(\ref{eq.chiral-ineq})   }
is chosen to derive bounds on
LECs up to $O(p^4)$ level. Thanks to a  numerical analysis,
we extract the most stringent bound in the region ${\cal R}$.
We have adopted  the input values $m_N=0.939$~GeV, $M_\pi=0.139$~GeV, $g_A=1.267$ and
$F_\pi=0.0924$~GeV, same as in~\cite{yao}.

The leading $O(p)$ pion-nucleon scattering amplitude is linear in $\nu$ and hence vanishes
when performing the second derivatives. Up to $O(p^2)$,
{ Eq.~(\ref{eq.chiral-ineq}) gives  for $c_2$ the bound}
\bea
f(\alpha,\nu,t)\,\,\,=\,\,\, 4\alpha \, c_2\,\,\, \geq \,\,\, 0\ .
\eea
Since $0.85\leq \alpha_{\rm min}(t)\leq \alpha\leq \alpha_{\rm max}(t)$
for $0\leq t\leq 4M_\pi^2$, the above inequality is simplified to $c_2\geq 0$.
It is trivial and well satisfied by {the} fit values
{$c_2=3.74\pm0.09$~GeV$^{-1}$  }    from Ref.~\cite{oller}  and
{$c_2=4.01\pm0.09$~GeV$^{-1}$   }   from Ref.~\cite{yao}
(see Table~\ref{tabwodel}).

\subsection{Analysis at $O(p^3)$ level}
\begin{figure*}[!t]
\centering
\begin{minipage}{0.4\textwidth}
\centering
\includegraphics[clip,width=1.0\textwidth]{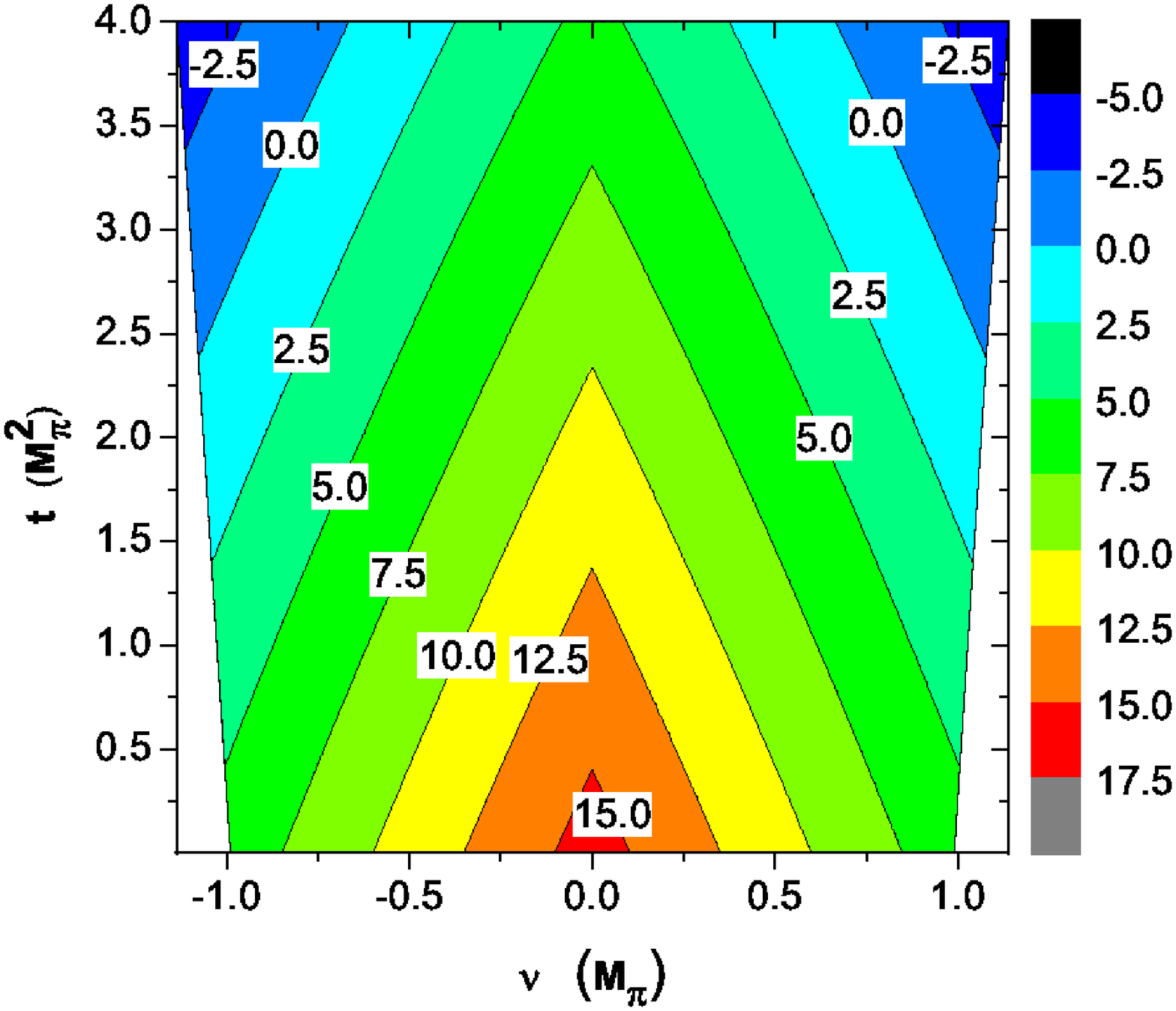}
\end{minipage}
\begin{minipage}{0.4\textwidth}
\centering
\includegraphics[clip,width=1.0\textwidth]{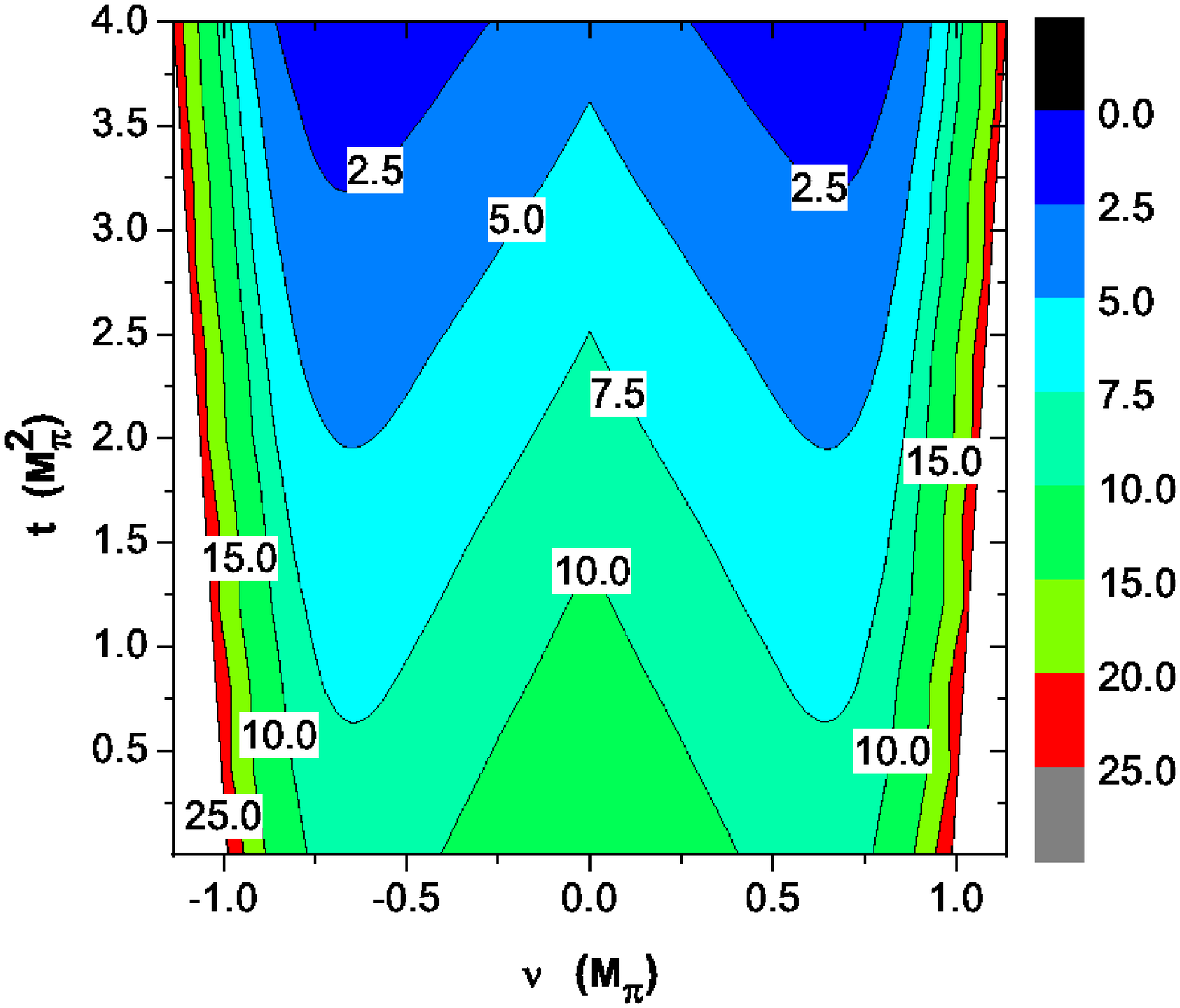}
\end{minipage}
\caption{\small \label{p3fitI}
{Positivity bound on LECs at $O(p^3)$ level. The fit results from `Fit I-$O(p^3)$' given
in Ref.~\cite{yao} are employed for plotting $f(\alpha,\nu,t)$ at $\cO(p^3)$.
Left-hand side: only tree-level; right-hand-side: tree-level + loops.
Similar results are obtained if one uses instead the  `WI08'  results with $\slashed{\Delta}$-ChPT
given in Ref.~\cite{oller}.
}
}
\end{figure*}

%%%\begin{figure*}[ht]
%%%\centering
%%%\begin{minipage}{0.4\textwidth}
%%%\centering
%%%\includegraphics[clip,width=1.0\textwidth]{p3bound1treeWI08.eps}
%%%\end{minipage}
%%%\begin{minipage}{0.4\textwidth}
%%%\centering
%%%\includegraphics[clip,width=1.0\textwidth]{p3bound1fullWI08.eps}
%%%\end{minipage}
%%%\caption{\small \label{p3WI08}Positivity bound on LECs at $O(p^3)$ level.
%%%The fit results from `WI08' of $\slashed{\Delta}$-ChPT  given in Ref.~\cite{oller} are employed for plotting. Left: tree, right: tree+loop}
%%%\end{figure*}

The scattering amplitudes in EOMS-B$\chi$PT were computed up to $O(p^3)$ in Refs.~\cite{oller}
and~\cite{yao} independently.  Therein, the amplitudes were employed to perform
fits to existing experimental phase-shift data , determining
the concerning LECs. Here, the positivity constraints, displayed by Eq.~(\ref{consonampl}),
provide additional information
about the amplitudes. When $n=2$, they  turn into Eq.~(\ref{eq.chiral-ineq})
and  give bounds on the LECs at  the $O(p^3)$ level:
\bea\label{boundp3}
f(\alpha,\nu,t) \quad =\quad
4\alpha c_2-8(\alpha-1)m_N(d_{14}-d_{15})-h_+^{(2)}(\alpha,\nu,t)
-\left|24\alpha\nu d_3-h_-^{(2)}(\alpha,\nu,t)\right|\geq0\ ,
\qquad\qquad   (\nu,t)\in{\cal R}\, ,
\eea
with   {  $\alpha_{\rm min}(t)\leq \alpha\leq \alpha_{\rm max}(t)$     }
and the second derivatives of
the non-pole loop contributions,
\bea
h_\pm^{(2)}(\alpha,\nu,t)=-F_\pi^2\frac{{\rm d}^2 \tilde{D}^{\pm,{\rm loop}}_\alpha(\nu,t)}{{\rm d}\nu^2}\ .
\eea
Note that the left-hand side of Eq.~(\ref{boundp3}) is a multivariate function
with respect to $\alpha$, $\nu$ and $t$.
%%%
%%%which will be denoted by $f(\alpha,\nu,t)$ hereafter.
%%%

The inequality given by Eq.~(\ref{boundp3}) is useful for judging the goodness  of the fit results
in Refs.~\cite{oller} and~\cite{yao}.  In both of them
the minimal value of $f(\alpha,\nu,t)$ is always achieved for $\alpha=\alpha_{\rm min}(t)$.
After  setting $\alpha=\alpha_{\rm min}(t)$, the  scanning of $(\nu,t)$ within the region ${\cal R}$
yields the most stringent bound for  $\nu=\pm0.68M_\pi,~t=4M_\pi^2$ in the $\cO(p^3)$ analysis~\cite{yao},
which is well respected: $f(\alpha_{\rm min}(4M_\pi^2),\pm0.68M_\pi,4M_\pi^2)=0.83\geq0$.
In a similar way, the $\cO(p^3)$ analysis~\cite{oller} produces
its most stringent bound for $\nu=\pm0.65M_\pi,~t=4M_\pi^2$ and $\alpha=\alpha_{\rm min}(4M_\pi^2)$,
which is well fulfilled:
$f(\alpha_{\rm min}(4M_\pi^2),\pm0.65M_\pi,4M_\pi^2)=1.03\geq0${.}

The contour plot for $f(\alpha,\nu,t)$ in the region ${\cal R}$,
with $\alpha=\alpha_{\rm min}(t)$,   is shown in Fig.~\ref{p3fitI}.
{We used the LEC central values from Ref.~\cite{yao}.}
We noticed that at $\cO(p^3)$ the EOMS-scheme renormalized loop contributions were numerically
relevant. If only the tree diagrams were considered in the inequality~(\ref{eq.chiral-ineq})
the corresponding bound fails  in some regions of $\mR$, where $f(\alpha,\nu,t)<0$
(see the left-hand side graph in Fig.~\ref{p3fitI}).
Hence, the loop contribution is crucial. It is needed not only at the formal level for the consistence
of the effective theory  but also for the numerical fulfillment of the positivity
 constraints at this chiral order.

The analyses above were carried out with the central values of the LECs.
In order to study the influence of the error and to provide a convenient inequality that can be used
in future analysis,
we take the particular point $\nu=\pm0.68M_\pi,~t=4M_\pi^2,~\alpha=\alpha_{\rm min}(4 M_\pi^2)=0.85$,
where the bound reads
\bea
f(\alpha,\nu,t)\quad =\quad
3.40c_2+1.11(d_{14}-d_{15})-0.29-|1.93d_3+1.22|\geq0\ ,\qquad\qquad
(\nu=\pm0.68M_\pi,~t=4M_\pi^2,~\alpha=0.85)\ ,
\label{eq.optimal-Op3}
\eea
{with $c_2$ and the $d_j$ given in GeV$^{-1}$ and GeV$^{-2}$ units, respectively. }
Notice that the numerical coefficients in this equation do not depend on $\cO(p^2)$
or $\cO(p^3)$ LECs, and are fully determined by $m_N$, $M_\pi$, $g_A$ and $F_\pi$.
Eq.~(\ref{eq.optimal-Op3}) provides the optimal bound for Ref.~\cite{yao} and nearly the optimal
for Ref.~\cite{oller}.
Considering now  the  $\cO(p^2)$ and $\cO(p^3)$ LEC uncertainties in the previous $\cO(p^3)$ inequality
one gets         {(in units of GeV$^{-1}$)}
\bear
f(\alpha,\nu,t) &=& 5.42\pm1.22-|-4.59\pm0.98|\overset{?}{\geq}0\qquad \mbox{(Ref.~\cite{yao}), }
\\
f(\alpha,\nu,t) &=& 4.89\pm1.23-|-3.84\pm0.98|\overset{?}{\geq}0\qquad  \mbox{(Ref.~\cite{oller}).}
\eear
See Table~\ref{tabwodel} for details on the LECs~\cite{oller,yao}. {Here the formula $\Delta f=\sqrt{\sum_i\left[f'(\overline{x}_i)\Delta x_i\right]^2}$ is adopted to propagate the errors of the LECs, where $x_i$ stands for the LECs with $\overline{x}_i$ the central values and $\Delta x_i$ the corresponding errors.} These expressions show a  violation of the positivity constrains
in part of the confidence region and queries the {convergence of} the pion-nucleon
scattering amplitude at the $O(p^3)$ level. Actually, this was  first pointed out by Ref.~\cite{oller}
where it was argued that the pion-nucleon calculation in EOMS scheme
may have problems with the convergence of the chiral expansion. This is partly confirmed  by the
$\cO(p^3)$ positivity analysis shown here, where not all the values  within the $1\sigma$ confidence
intervals fulfill the bound.
Thus, the constraint~(\ref{eq.optimal-Op3})  may help to stabilize future fits to data
and the chiral expansion.

\begin{table}[!t]
\begin{center}
\caption{\label{tabwodel} LECs involved in the positivity bounds without
{explicit  $\Delta$(1232) contributions.}
Actually, the $c_i$ in the Fit I(a)-$O(p^4)$~\cite{yao} stand for $\hat{c}_i$. The $*$ denotes an input
quantity. The $c_i$, $d_j$
and $e_k$ have units of GeV$^{-1}$,
GeV$^{-2}$ and GeV$^{-3}$,   respectively. }
\begin{tabular}{c c c c c c c}
\hline\hline
 LEC    &        Fit I-$O(p^3)$~\cite{yao}      & WI08 ($\slashed{\Delta}$-ChPT)~\cite{oller} & Fit I(a)-$O(p^4)$~\cite{yao} & Fit I(b)-$O(p^4)$~\cite{yao} & Fit I(c)-$O(p^4)$~\cite{yao}\\
\hline
$c_1$   &       $-1.39\pm0.07$     & $-1.50\pm0.06$   & $-1.08\pm0.06$  & $-1.39^*$      &$-1.09\pm0.08$\\
$c_2$   &       $4.01\pm0.09 $      & $3.74\pm0.09$    & $2.78\pm0.11$   & $4.01^*$      &$2.24\pm0.05$\\
$c_3$   &       $-6.61\pm0.08$     & $-6.63\pm0.08$   & $-5.26\pm0.14$  & $-6.61^*$      &$-5.05\pm0.22$\\
$c_4$   &       $3.92\pm0.04 $      & $3.68\pm0.05$    & $2.43\pm0.19$   & $3.92^*$      &$2.43\pm0.19$\\
$d_3$   &       $-3.02\pm0.51$     & $-2.63\pm0.51$   & $-6.87\pm0.16$  & $-8.04\pm0.13$ &$-6.87\pm0.15$\\
$d_{14}-d_{15}$&$-7.15\pm1.06$      &$-6.80\pm1.07$    & $-12.09\pm0.24$   & $-13.90\pm0.20$          &$-11.94\pm0.23$\\
$e_{15}$&        $\cdots$                &  $\cdots$            & $-14.99\pm0.55$   &$-14.50\pm0.55$  &$-5.41\pm0.57$\\
$e_{16}$   &     $\cdots$           &  $\cdots$           & $7.35\pm0.35$  &$7.65\pm0.35$             &$4.34\pm0.28$\\
$e_{18}$   &     $\cdots$           & $\cdots$          & $6.07\pm1.18$  &$-0.79\pm1.19$              &$6.00\pm1.26$\\
$e_{20}+e_{35}$& $\cdots$           &  $\cdots$           & $\cdots$          &$-12.86\pm0.83$        &$\cdots$ \\
$e_{22}-4e_{38}$& $\cdots$             & $\cdots$         & $\cdots$         &$-8.19\pm1.79$          &$\cdots$ \\
\hline\hline
\end{tabular}
\end{center}
\end{table}

\subsection{Analysis at $O(p^4)$ level}

\begin{figure*}[!t]
\centering
\begin{minipage}{0.4\textwidth}
\centering
\includegraphics[clip,width=1.0\textwidth]{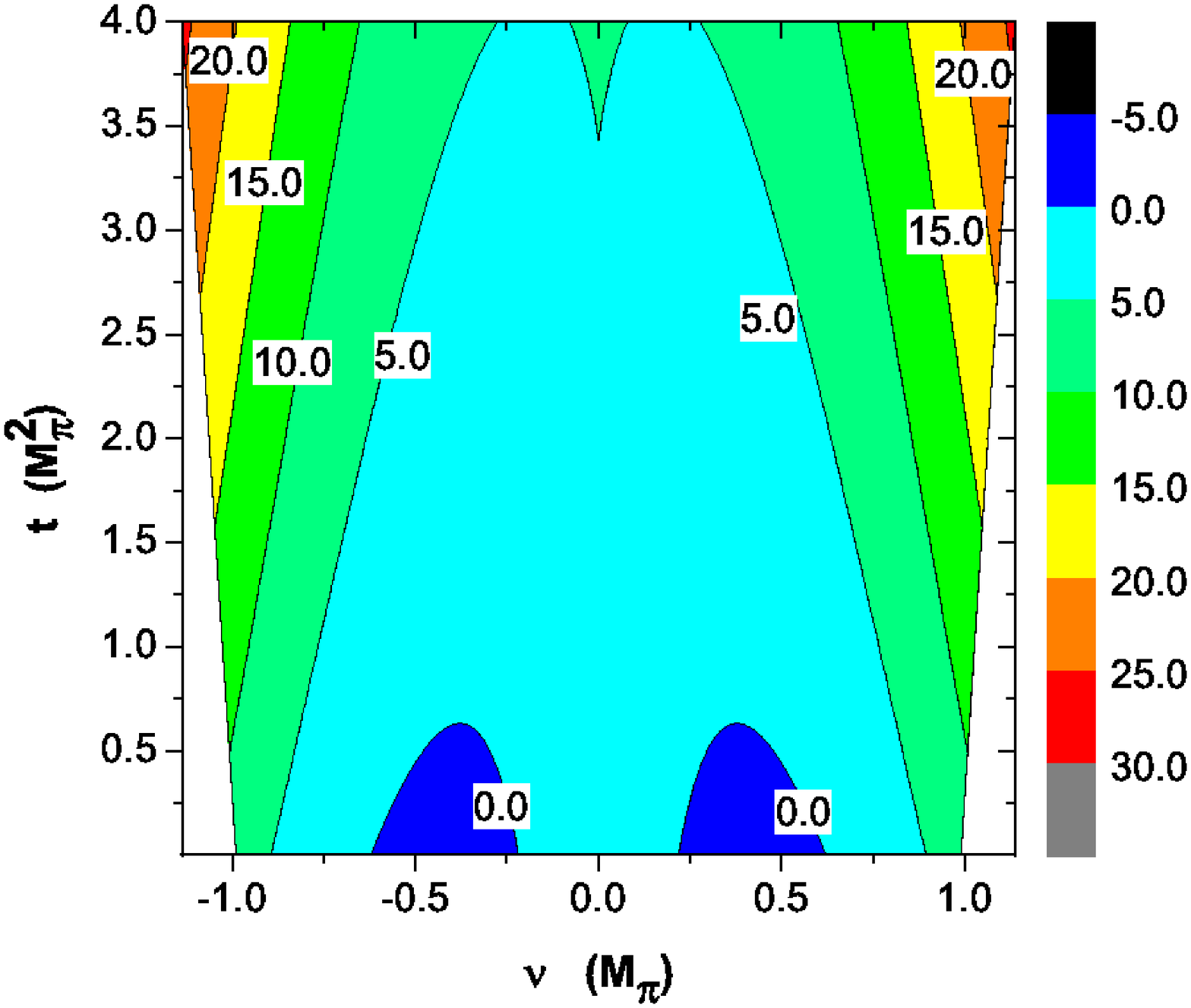}
\end{minipage}
\begin{minipage}{0.4\textwidth}
\centering
\includegraphics[clip,width=1.0\textwidth]{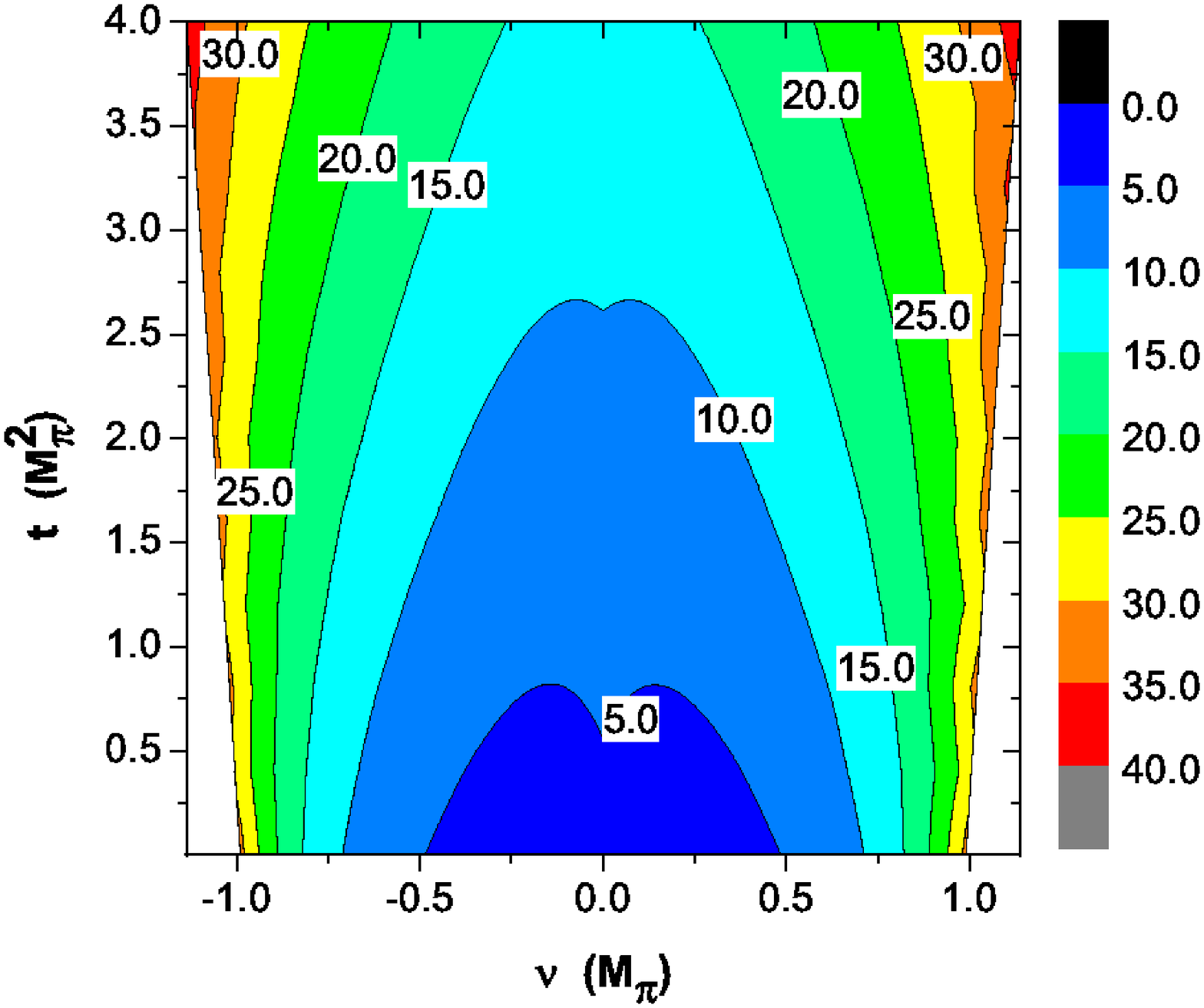}
\end{minipage}
\caption{\label{p4fitIa}\small
Positivity bound on LECs at $O(p^4)$ level. The fit results from
{`Fit I(a)-$O(p^4)$'}
given in Ref.~\cite{yao} are employed for plotting
{$f(\alpha,\nu,t)$ up to $\cO(p^4)$ in EOMS-B$\chi$PT.}
{\bf Left-hand side:} only tree-level; {\bf right-hand side:} tree+loop}
\end{figure*}

\begin{figure*}[!t]
\centering
\begin{minipage}{0.4\textwidth}
\centering
\includegraphics[clip,width=1.0\textwidth]{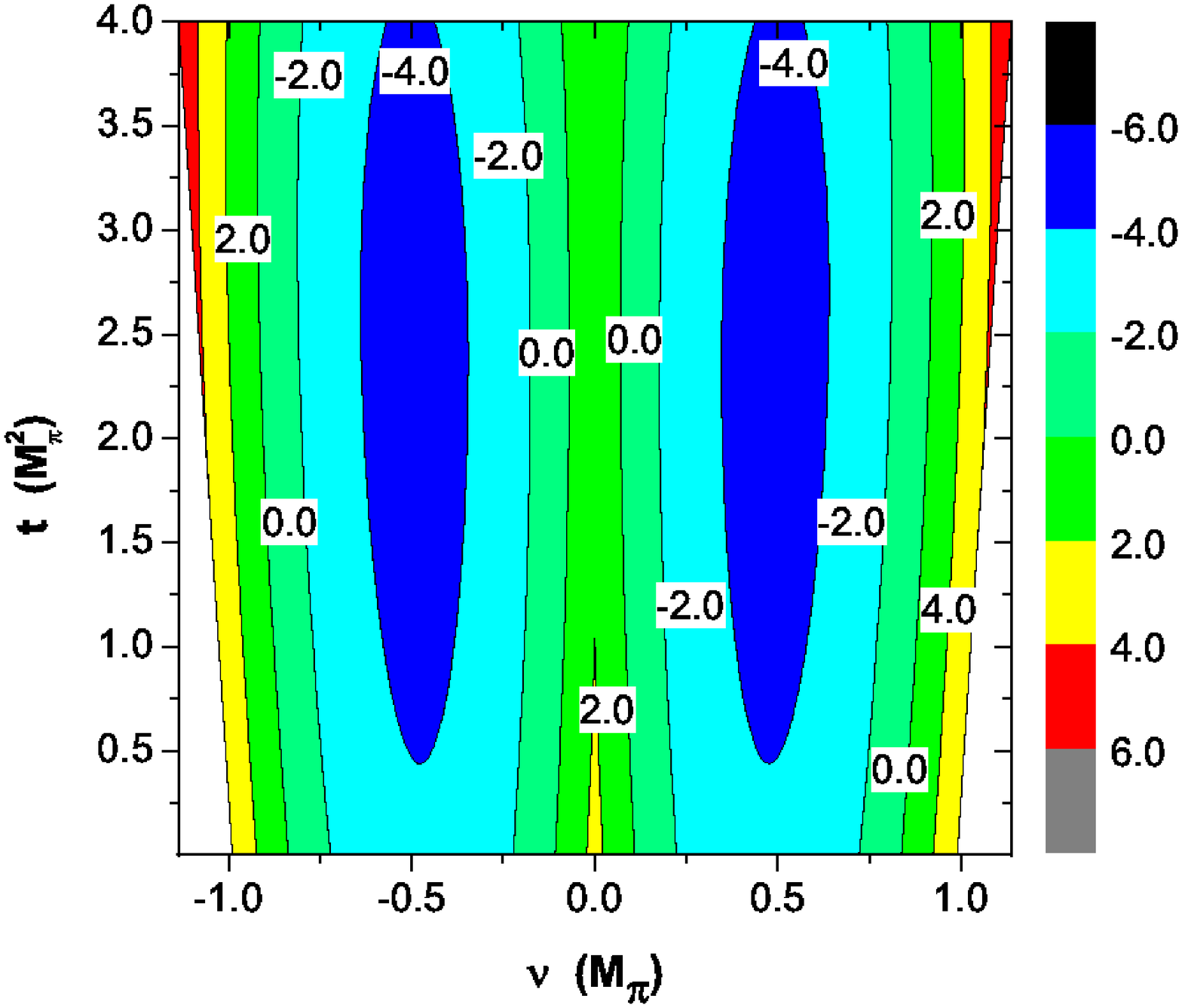}
\end{minipage}
\begin{minipage}{0.4\textwidth}
\centering
\includegraphics[clip,width=1.0\textwidth]{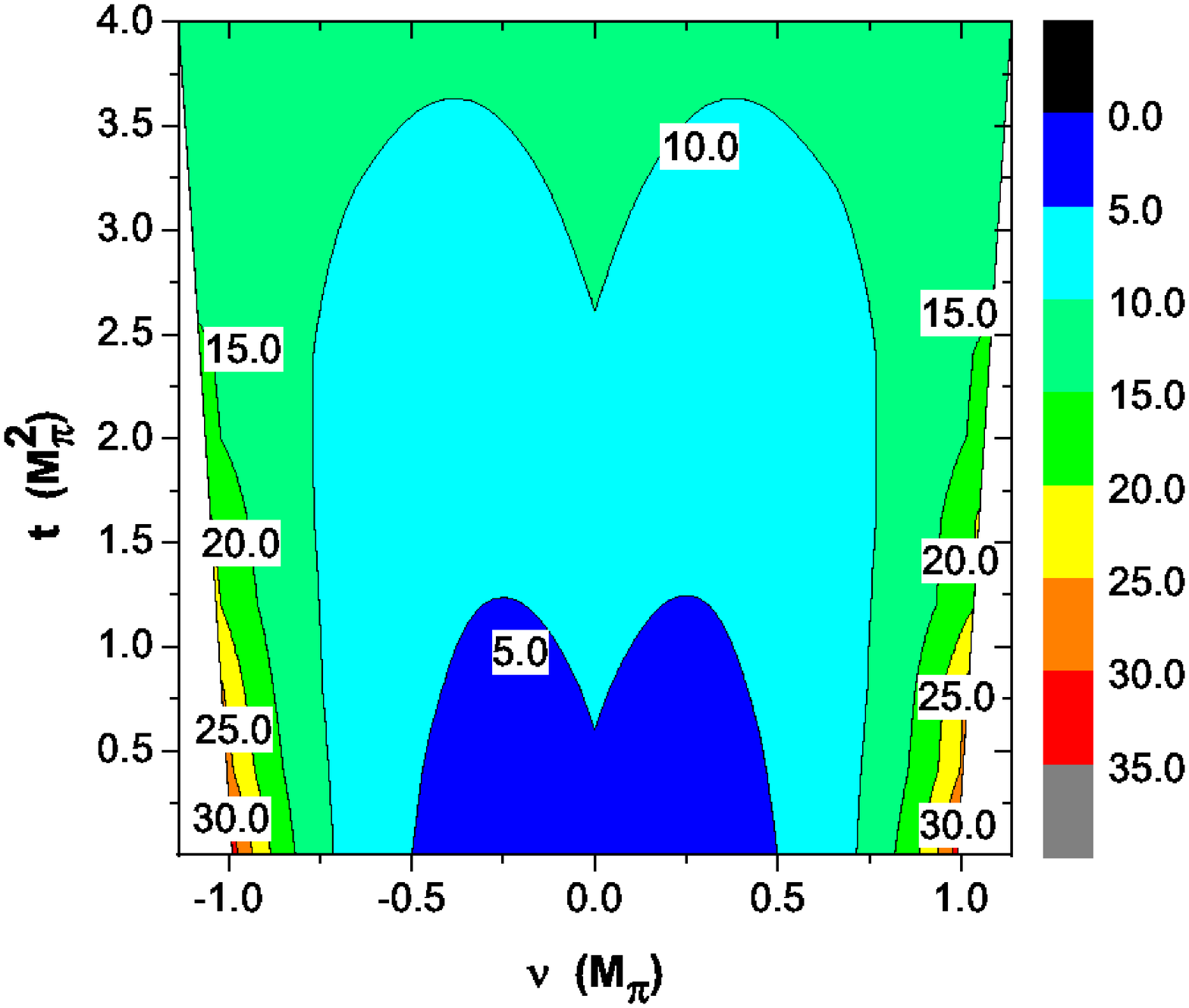}
\end{minipage}
\caption{\label{p4fitIb}\small
Positivity bound on LECs at $O(p^4)$ level. The fit results from `Fit I(b)-$O(p^4)$'
given in Ref.~\cite{yao} are employed for plotting
{$f(\alpha,\nu,t)$ up to $\cO(p^4)$ in EOMS-B$\chi$PT.}
{\bf Left-hand side:} only tree-level;  {\bf right-hand side:} tree+loop}
\end{figure*}

At the $O(p^4)$ level, {with two subtractions ($n=2$), }
the bound~(\ref{eq.chiral-ineq})
on the LECs turns into
\bea\label{boundp4a}
&&f(\alpha,\nu,t) \quad =\quad 4\alpha \hat{c}_2-8(\alpha-1)m_N(d_{14}-d_{15})
+32\alpha\left[-\hat{c}_1\hat{c}_2M_\pi^2-2e_{15}m_N\nu_B
+6e_{16}\nu^2\right]-h_+^{(2)}(\alpha,\nu,t)\nonumber\\
&&\qquad \qquad\qquad\qquad\qquad
-\left|24\alpha\nu d_3-96(\alpha-1)e_{18}m_N\nu-h_-^{(2)}(\alpha,\nu,t)\right|\geq0\ ,
\eea
where $\hat{c}_1=c_1-2M_\pi^2(e_{22}-4e_{38})$ and $\hat{c}_2=c_2+8M_\pi^2(e_{20}+e_{35})$~\cite{yao}.
Here the non-pole loop terms $h_\pm^{(2)}(\alpha,\nu,t)$
contain both $O(p^3)$ and $O(p^4)$ contributions.
It is useful to reexpress the bound~(\ref{boundp4a}) up to $\cO(p^4)$  as
\bea\label{boundp4b}
&&f(\alpha,\nu,t)\quad =\quad 4\alpha c_2-8(\alpha-1)m_N(d_{14}-d_{15})+32\alpha\left[(e_{20}+e_{35}-c_1c_2)M_\pi^2-2e_{15}m_N\nu_B
+6e_{16}\nu^2\right]-h_+^{(2)}(\alpha,\nu,t)
\nonumber\\
&&\qquad\qquad\qquad\qquad\qquad
-\left|24\alpha\nu d_3-96(\alpha-1)e_{18}m_N\nu-h_-^{(2)}(\alpha,\nu,t)\right|\geq0\ .
\eea
These two  equations differ from each other by terms of $\cO(p^5)$ in the chiral expansion or higher.

Two different strategies were adopted in Ref.~\cite{yao}
to perform fits to the pion-nucleon phase-shift and to determine the various LECs at $O(p^4)$
level within EOMS-B$\chi$PT. The strategy called `Fit I(a)-$O(p^4)$' provides values
for the LECs in Eq.~(\ref{boundp4a}), and the other one, called `Fit I(b)-$O(p^4)$',
gives values for the LECs in Eq.~(\ref{boundp4b}).
%%%
%%%In this paper, the aim is to use Eq.~\ref{boundp4a} and Eq.~\ref{boundp4b} to test those fit
%%%results from `Fit I(a)-$O(p^4)$' and `Fit I(b)-$O(p^4)$', respectively.
%%%
As it happened before at $\cO(p^3)$, the function $f(\alpha,\nu,t)$ up to $\cO(p^4)$
also achieves its minimal values  for $\alpha=\alpha_{\rm min}(t)$.
For the central values of the LECs (see Table~\ref{tabwodel}), the $\cO(p^4)$ contour plot
for $f(\alpha,\nu,t)$ in the region ${\cal R}$, with $\alpha=\alpha_{\rm min}(t)$,
are shown in Figs.~\ref{p4fitIa} and~\ref{p4fitIb} .These two figures
correspond  to the two different $\cO(p^4)$  analysis,
'Fit I(a)-$\cO(p^4)$'~\cite{yao} in Eq.~(\ref{boundp4a}) and
'Fit I(b)-$\cO(p^4)$'~\cite{yao} in Eq.~(\ref{boundp4b}), respectively.
The most stringent bound stemming from  Eq.~(\ref{boundp4a})  takes the form
\bea\label{boundp4aex}
&&f(\alpha,\nu,t)\quad =\quad
4.36\hat{c}_2\,\,\,-\, \,\, 2.74-0.23\hat{c}_1-0.23\hat{c}_3+0.14\hat{c}_4
+0.11e_{16}+0.62(e_{15}-\hat{c}_1\hat{c}_2)
\nonumber\\
&&\qquad\qquad \qquad \qquad
-|0.18-0.26\hat{c}_1-0.07\hat{c}_2-0.24\hat{c}_3+0.20\hat{c}_4+0.57d_3|
\quad \geq \quad 0\ ,
\qquad \quad  (\nu=\pm0.17M_\pi,~t=0,~\alpha=\alpha_{\rm min})\ ,
\nonumber\\
\eea
{with the $c_i$, $d_j$ and $e_k$ in units of GeV$^{-1}$, GeV$^{-2}$
and GeV$^{-3}$, respectively.   }
Substituting the LECs in Eq.~(\ref{boundp4aex}) with the values from `Fit I(a)-$O(p^4)$'~\cite{yao}
(see Table~\ref{tabwodel}),  one finds          {(in units of GeV$^{-1}$)}
\bear
f(\alpha,\nu,t) &=& 4.56\pm0.66-|-1.88\pm0.11|\geq0 \, ,
\eear
{where}   the positivity constraint is definitely well obeyed at the $O(p^4)$ level in the chiral expansion.

In the alternative $\cO(p^4)$ form~(\ref{boundp4b}) the most stringent bound reads
\bea
&&f(\alpha,\nu,t)\quad=\quad
4.36c_2 \,\,\, -\,\,\, 2.72-0.23c_1-0.23c_3+0.14c_4+0.11e_{16}+0.62(e_{15}+e_{20}+e_{35}-c_1c_2)
\nonumber\\
&&\qquad\qquad\qquad\qquad
-|0.19-0.27c_1-0.08c_2-0.26c_3+0.21c_4+0.60d_3|\geq0\ ,\qquad \qquad
(\nu=\pm0.18M_\pi,~t=0,~\alpha=\alpha_{\rm min})\ ,
\nonumber\\
\eea
{with the $c_i$, $d_j$ and $e_k$ in units of GeV$^{-1}$, GeV$^{-2}$
and GeV$^{-3}$, respectively.   }
Substituting the values from `Fit I(b)-$O(p^4)$'~\cite{yao} in Table~\ref{tabwodel},
we find that  the positivity bound  is again well respected at   $O(p^4)$:
{\bear
f(\alpha,\nu,t) &=&  4.61\pm0.62-|-2.04\pm0.08|\geq0\, ,
\eear }
{given in units of GeV$^{-1}$.}

As it happened at $\cO(p^3)$, our $\cO(p^4)$ analyses in Figs.~\ref{p4fitIa} and \ref{p4fitIb}
(left-hand side)
show that in the EOMS scheme the bounds are violated in some regions of $\mR$
if only the (renormalized) tree-level amplitude is included; loops play an important
role, both at $O(p^3)$ and $O(p^4)$.

\subsection{Comparison at special subthreshold points}

At the subthreshold region, some famous low-energy theorems can be established at particular points:
the Cheng-Dashen (CD) point $(\nu=0,\, t=2 M_\pi^2)$~\cite{CDpoint} and the Adler point
$(\nu=0,\, t= M_\pi^2$)~\cite{Adlerpoint}.
The  positivity bound is found to be very clearly obeyed at these points, both at $O(p^3)$ and $O(p^4)$
(see Figs~\ref{p3fitI}-\ref{p4fitIb}).
Nonetheless , it is still interesting to study the evolution of the constraints at these points
as the chiral order increases from $O(p^3)$ to $O(p^4)$.
{\it A priori}, the variation of the bounds at the CD  and Adler points should not be too large,
since  the chiral convergence of the amplitudes is expected to be good ($\nu \ll m_N$ and $t\ll m_N^2$) and
these points are far away from non-analytical points.
On the other hand, the bounds near threshold always get large values for $f(\alpha,\nu,t)$
and suffer   {a}   sizable variation  from one chiral order to another as  the derivatives of the loop amplitude
may diverge at threshold.
In what follows, the bounds at these special subthreshold points will be calculated with the condition
$\alpha=\alpha_{\rm min}(t)$,
where we extracted the most stringent bounds in the  sections above.

At the CD point, where $(\nu=0,\,t=2M_\pi^2)$,
setting $\alpha = \alpha_{\rm min}(2M_\pi^2)$
%%%
%%%\bea
%%%1\,\,\, -\,\,\, \frac{M_\pi}{2m_N}
%%%\,\,\, -\,\,\, \frac{M_\pi}{2m_N}
%%%\,\,\, -\,\,\, \frac{M_\pi^2}{2m_N^2}
%%%\,\,\, +\,\,\, \cO\left(\Frac{M_\pi^3}{m_N^3}\right) \,\,\,
%%%\simeq \,\,\, 0.917\, .
%%%%%%
%%%%%%\cong0.926\ .   <<<<<<<<<<<<<<\cO(p^3)
%%%%%%
%%%\eea
%%%
the $O(p^3)$ bound~(\ref{boundp3}) reads          {(in units of GeV$^{-1}$)}
\bea
f(\alpha_{\rm min}(2M_\pi^2),0,2M_\pi^2)^{\cO(p^3)}\quad = \quad \left\{
                                 \begin{array}{ll}
                                   8.7\, ,         %%%8.68,
                                   &\quad  \hbox{for `Fit I-$O(p^3)$' from Ref.~\cite{yao},} \\
                                   7.9\, ,          %%%7.88,
                                   & \quad \hbox{for `WI08' of $\slashed{\Delta}$-ChPT from Ref.~\cite{oller},}
                                 \end{array}
                               \right.
\label{eq.CD-op3}
\eea
and the $O(p^4)$ bounds~(\ref{boundp4a}) and (\ref{boundp4b})  become now          {(in units of GeV$^{-1}$)}
\bea
f(\alpha_{\rm min}(2M_\pi^2),0,2M_\pi^2)^{\cO(p^4)}=\left\{
                                 \begin{array}{ll}
                                   8.5\, ,           %%%9.16,
                                   & \quad \hbox{for `Fit I(a)-$O(p^4)$' from~\cite{yao},} \\
                                   8.5\, ,            %%%9.09,
                                   & \hbox{for `Fit I(b)-$O(p^4)$' from~\cite{yao}.}
                                 \end{array}
                               \right.
\eea
As expected, the $O(p^3)$ bounds at the CD point, located at the center of the upper part $\mR$ of the Mandelstam triangle,
suffer small variations when taking the EOMS-B$\chi$PT up to  $O(p^4)$.

In a similar way,
{taking the optimal value  $\alpha=\alpha_{\rm min}(M_\pi^2)$},
the  $O(p^3)$ bound~(\ref{boundp3})  at the Adler point  $(\nu=0,\, t=M_\pi^2)$
reads          {(in units of GeV$^{-1}$)}
\bea
f(\alpha_{\rm min}(M_\pi^2),0,M_\pi^2)^{\cO(p^3)} \quad =\quad  \left\{
                                 \begin{array}{ll}
                                   11.0\, ,                   %%%10.92,
                                   & \quad \hbox{for `Fit I-$O(p^3)$' from Ref.~\cite{yao},} \\
                                   10.0\, ,                   %%%9.97,
                                   & \hbox{for `WI08' of $\slashed{\Delta}$-ChPT from Ref.~\cite{oller},}
                                 \end{array}
                               \right.
\eea
and the $O(p^4)$ bounds~(\ref{boundp4a}) and (\ref{boundp4b})
give          {(in units of GeV$^{-1}$)}
\bea
f(\alpha_{\rm min}(M_\pi^2),0,M_\pi^2)^{\cO(p^4)} \quad =\quad \left\{
                                 \begin{array}{ll}
                                   6.1\, ,                    %%%6.52,
                                   & \hbox{for `Fit I(a)-$O(p^4)$' from~\cite{yao},} \\
                                   6.0\, ,                    %%%6.42,
                                   & \hbox{for `Fit I(b)-$O(p^4)$' from~\cite{yao}.}
                                 \end{array}
                               \right.
\eea
Compared  to the CD point, the variation of the bound at the Adler point is slightly  larger,
{yet}   still  rather acceptable.

\subsection{Analysis including the $\Delta(1232)$\label{secdel}}

\begin{table}[!t]
\begin{center}
\caption{\label{tabwdel}\small
LECs involved in the positivity bounds with {explicit}   $\Delta$(1232) contribution. The $*$ denotes an input
quantity. The $c_i'$, $d_j'$
and $e_k'$ have units of GeV$^{-1}$,
GeV$^{-2}$ and GeV$^{-3}$ respectively, and $h_A$ is dimensionless.}
\begin{tabular}{c c c c c c c}
\hline\hline
 LEC    &        Fit II-$O(p^3)$~\cite{yao}    & WI08 ($\Delta$-ChPT)~\cite{oller} & Fit II(a)-$O(p^4)$~\cite{yao} & Fit II(b)-$O(p^4)$~\cite{yao} & Fit II(c)-$O(p^4)$~\cite{yao}\\
\hline
$c_1'$   &       $-0.81\pm0.03$     & $-1.00\pm0.04$   & $-1.03\pm0.03$  & $-0.81^*$  &$-0.95\pm0.05$\\
$c_2'$   &       $1.46\pm0.09 $      & $1.01\pm0.04$    & $0.50\pm0.04$   & $1.46^*$  &$0.10\pm0.06$\\
$c_3'$   &       $-3.10\pm0.12$     & $-3.04\pm0.02$   & $-3.17\pm0.05$  & $-3.10^*$  &$-2.64\pm0.08$\\
$c_4'$   &       $2.35\pm0.06 $      & $2.02\pm0.01$    & $0.79\pm0.03$   & $2.35^*$  &$0.80\pm0.03$\\
$d_3'$   &       $-0.47\pm0.05$     & $-0.23\pm0.27$   & $-5.04\pm0.05$  & $-4.75\pm0.04$&$-4.90\pm0.04$\\
$d_{14}'-d_{15}'$&$-0.90\pm0.15$      &$-0.50\pm0.50$    & $-5.61\pm0.09$   & $-5.82\pm0.09$&$-5.58\pm0.09$\\
$e_{15}'$&       $\cdots$            &  $\cdots$           & $5.05\pm0.13$   &$15.29\pm0.12$&$10.52\pm0.12$\\
$e_{16}'$   &     $\cdots$            &  $\cdots$          & $-0.31\pm0.07$  &$-2.76\pm0.07$&$-1.50\pm0.05$\\
$e_{18}'$   &     $\cdots$           & $\cdots$           & $-10.99\pm0.12$  &$-11.58\pm0.11$&$-9.87\pm0.12$\\
$e_{20}'+e_{35}'$& $\cdots$             &  $\cdots$            & $\cdots$          &$-13.12\pm0.28$&  $\cdots$\\
$e_{22}'-4e_{38}'$& $\cdots$             & $\cdots$          & $\cdots$          &$10.29\pm0.82$&  $\cdots$\\
$h_A$   &       $2.82\pm0.04$      & $2.87\pm0.04$   & $2.90^*$  & $2.90^*$& $2.90^*$\\
\hline\hline
\end{tabular}
\end{center}
\end{table}

\begin{figure*}[!t]
\centering
\begin{minipage}{0.4\textwidth}
\centering
\includegraphics[clip,width=1.0\textwidth]{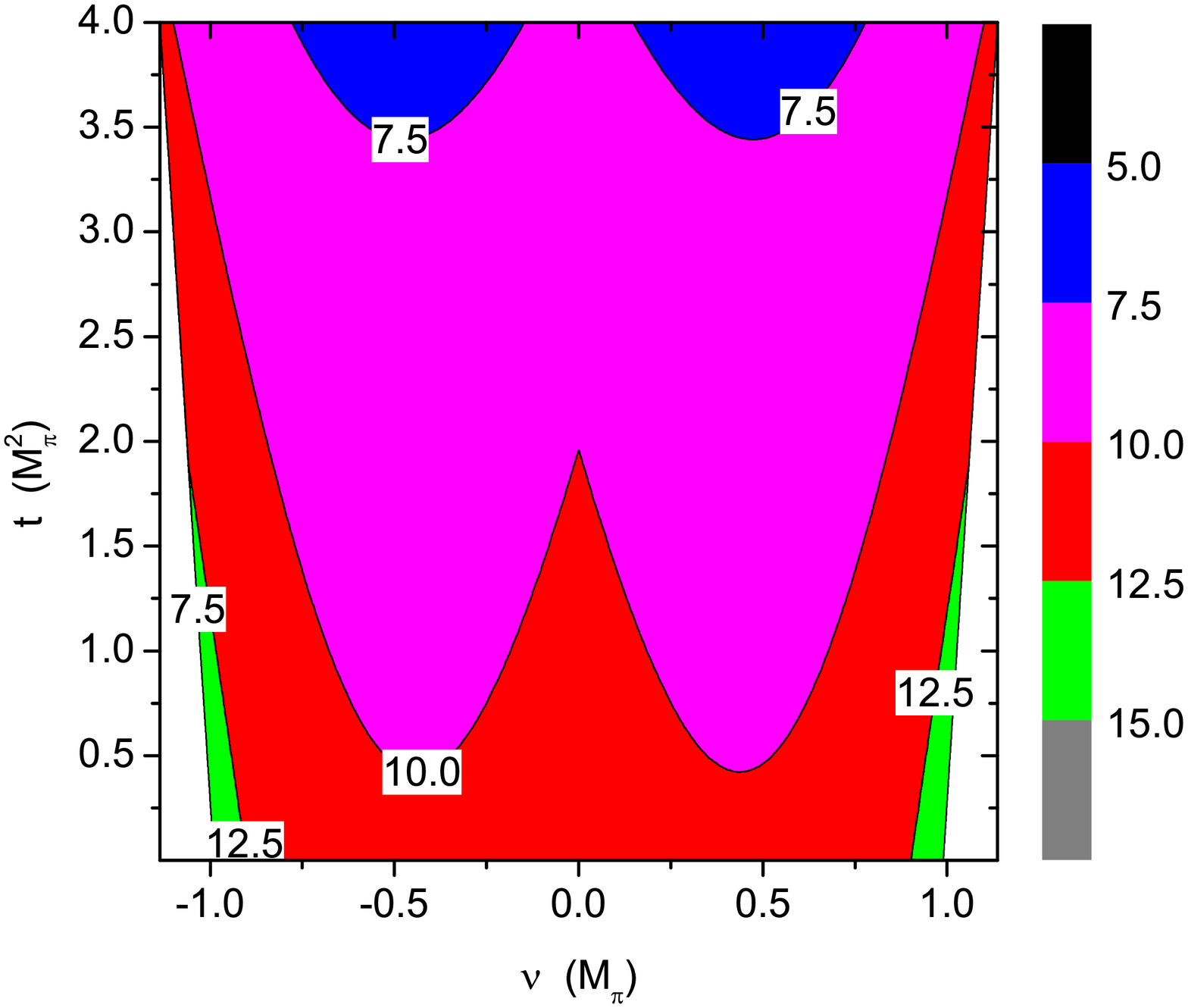}
\end{minipage}
\begin{minipage}{0.4\textwidth}
\centering
\includegraphics[clip,width=1.0\textwidth]{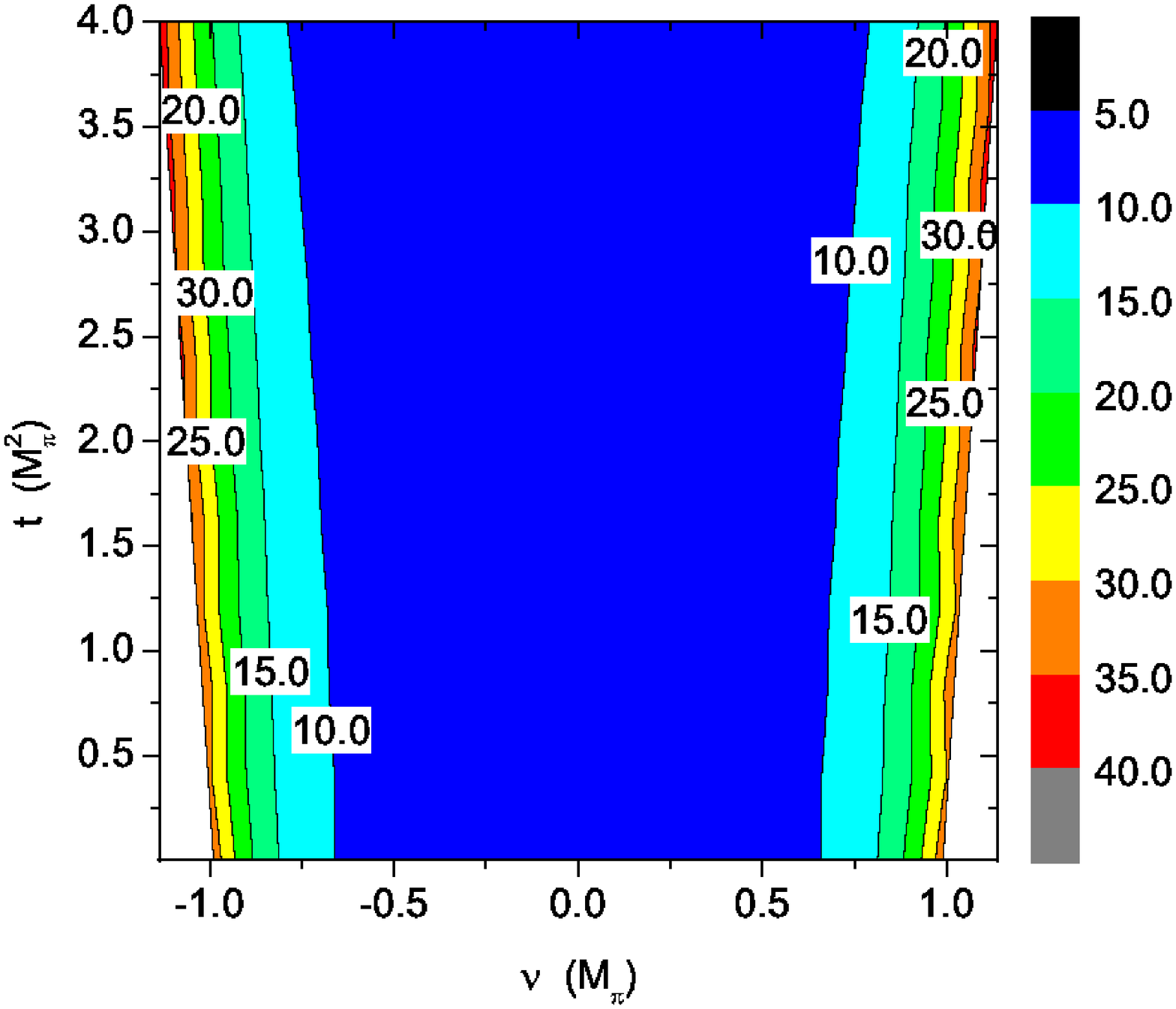}
\end{minipage}
\caption{\label{p3fitII}\small
Positivity bound on LECs at $O(p^3)$ level including the $\Delta(1232)$.
The fit results from `Fit II-$O(p^3)$' given in Ref.~\cite{yao} are employed for plotting $f(\alpha,\nu,t)$.
{\bf Left-hand side:} only tree-level; {\bf right-hand side:} tree+loop.
The analysis `WI08' with $\Delta$-ChPT  in Ref.~\cite{oller} yields a similar outcome.
}
\end{figure*}

%%%\begin{figure*}[ht]
%%%\centering
%%%\begin{minipage}{0.4\textwidth}
%%%\centering
%%%\includegraphics[clip,width=1.0\textwidth]{deltap3WI08tree.eps}
%%%\end{minipage}
%%%\begin{minipage}{0.4\textwidth}
%%%\centering
%%%\includegraphics[clip,width=1.0\textwidth]{deltap3WI08full.eps}
%%%\end{minipage}
%%%\caption{\label{p3WI08D}Positivity bound on LECs at $O(p^3)$ level. The fit results from `WI08' of $\Delta$-ChPT  given in Ref.~\cite{oller} are employed for plotting. Left: tree, right: tree+loop}
%%%\end{figure*}

{
In Refs.~\cite{oller,yao}, the contribution from the $\Delta(1232)$
was explicitly included to describe the phase-shift
up to center-of-mass energies of $1.20$~GeV. The corresponding LECs
were pinned down through  fits to the experimental data.
%%%
%%%Hence, similar to the analyses without the $\Delta(1232)$ given above,
%%%it is straightforward to make a parallel study to judge the validity of the fits including the $\Delta(1232)$.
%%%
the value of $f(\alpha,\nu,t)$ can be readily obtained from the EOMS-B$\chi$PT
bounds at $\cO(p^3)$
{  (Eq.~(\ref{boundp3}))   }
and $\cO(p^4)$ (Eqs.~(\ref{boundp4a}) and (\ref{boundp4b}))
by conveniently adding the corresponding contributions
$F_\pi^2 \frac{d^2}{d\nu^2} \tilde{D}^\pm_\alpha(\nu,t)|^{\rm \Delta-Born}$.
In addition, at $\cO(p^4)$ in the $\delta$ counting~\cite{Pascalutsa} one may have contributions from $\Delta$ resonance
loops and the $\cO(p^2)$ LECs in the one-loop diagrams need to be modified (see App.~A.2 in Ref.~\cite{yao}).
The contour plots for $f(\alpha,\nu,t)$  inside the upper part of the Mandelstam triangle for the $O(p^3)$ amplitude
including the $\Delta(1232)$ is  shown in Fig.~\ref{p3fitII}.        %%% and~\ref{p3WI08D}.
Here we provided the fit results from `Fit II-$O(p^3)$'~\cite{yao}.
The `WI08' analysis in Ref.\cite{oller} produces similar results.    %%%are employed for plotting Fig~\ref{p3fitII} and Fig~\ref{p3WI08D}, respectively.
The $\cO(p^3)$ calculations~\cite{oller,yao} took the  $\Delta$(1232) into consideration by adding the
leading $\Delta$--Born term contribution explicitly (see the Appendices therein).
We find that this leading $\Delta$--Born term provides a definite positive and large contribution
to the $O(p^3)$ bounds (see Fig.~\ref{p3fitII}),
and  both the tree-level and the full (tree+loop) bound are well obeyed.
}

\begin{figure*}[!t]
\centering
\begin{minipage}{0.4\textwidth}
\centering
\includegraphics[clip,width=1.0\textwidth]{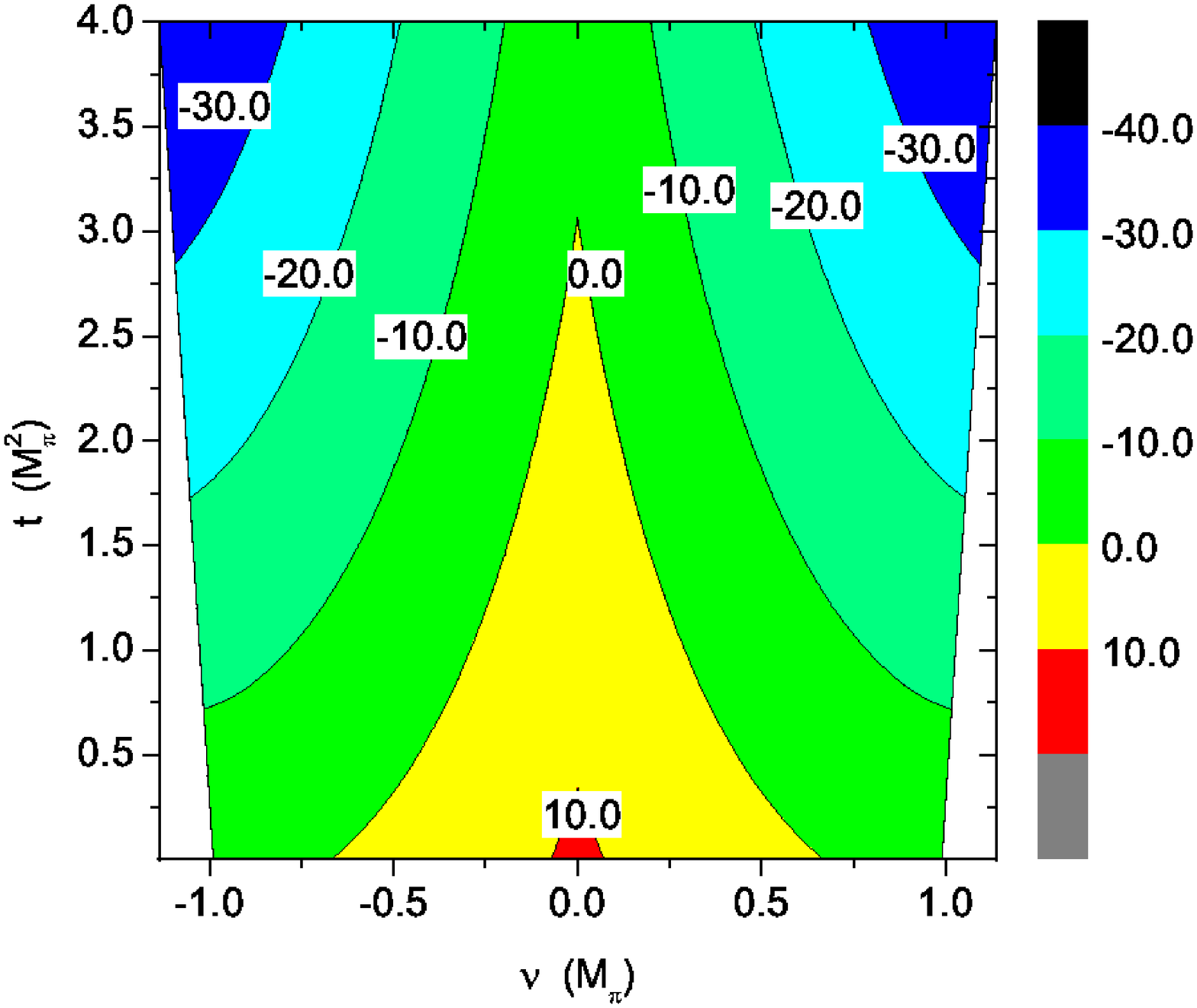}
\end{minipage}
\begin{minipage}{0.4\textwidth}
\centering
\includegraphics[clip,width=1.0\textwidth]{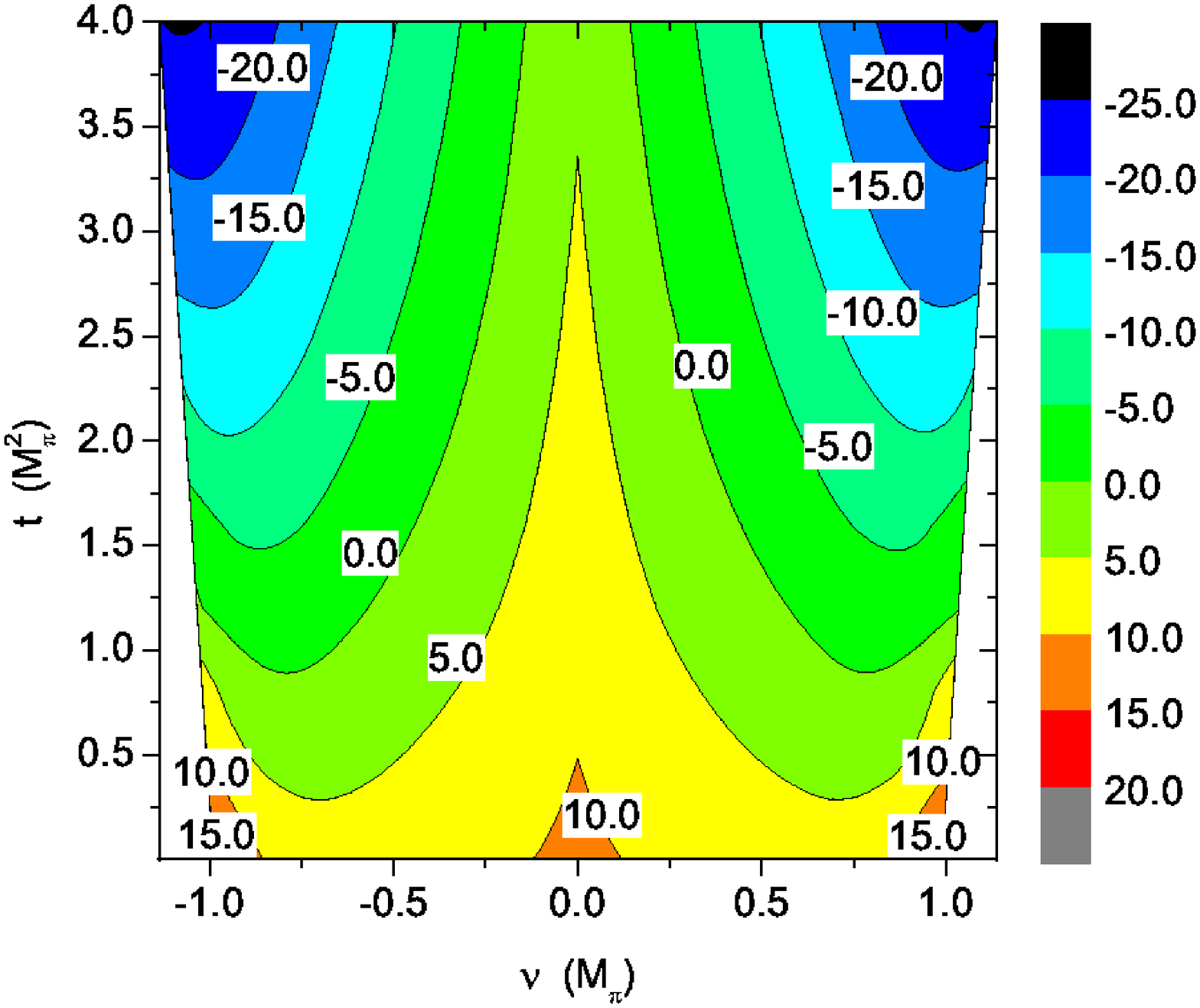}
\end{minipage}
\caption{\label{p4fitaII}\small
Positivity bound on LECs at $O(p^4)$ level. The fit results from `Fit II(a)-$O(p^4)$'
given in Ref.~\cite{yao} are employed for plotting $f(\alpha,\nu,t)$.
{\bf Left-hand side:} only tree-level; {\bf right-hand side:} tree+loop.
Similar results are found with `Fit II(b)-$O(p^4)$' from Ref.~\cite{yao}.
}
\end{figure*}

{
At $O(p^4)$,  the leading order Born contribution from explicit
$\Delta$(1232) exchanges were considered in Ref.~\cite{yao} and  the
$\Delta$(1232) loop contributions  were also partially included.
Therein, two scenarios were carried out, ``Fit II(a)''  and   ``Fit II(b)'',
corresponding to the two different ways of writing down the $\cO(p^4)$ part shown
in Eqs.~(\ref{boundp4a}) and (\ref{boundp4b}), but now including explicitly the $\Delta(1232)$.
At the $\cO(p^4)$ chiral order one needs to take into account the $\Delta$ resonance loops.
Their~$\cO(p^4)$ contribution was accounted in Ref.~\cite{yao} by adding the $\Delta$ contributions
$c_1^\Delta=0$, $c_2^\Delta = - c_3^\Delta = 2 c_4^\Delta = h_A^2 m_N^2/[9 m_\Delta^2 (m_\Delta -m_N)]$
to the $\cO(p^2)$ parameters $c_k$ present in the $\cO(p^4)$ B$\chi$PT loop .
Fig.~\ref{p4fitaII} shows the $f(\alpha,\nu,t)$ contour plot for ``Fit II(a)'', having
``Fit II(b)''   a similar structure.
%%%
%%%which explicitly includes $\Delta$(1232) contribution, are shown in Figs~\ref{p4fitaII}
%%%and~\ref{p4fitbII}, respectively.
%%%
The left-hand side  graph in Fig.~\ref{p4fitaII}           %%% and~\ref{p4fitbII}
presents the contour plots if only the tree-level amplitude is taken into account,
while the right-hand side shows the full bounds (tree+loop).
}

It is shocking that both the tree-level and full bounds are largely violated in the upper left and right
corners of the region $\mR$.
%%%
%%%, for both  ``Fit II(a)'' and ``Fit II(b)''.
%%%
The violation of the positivity bounds implies a possible issue
in the $O(p^4)$ fit results with the $\Delta$(1232) in Ref.~\cite{yao}.
To have a better understanding of this violation, one {should pay}
attention to the unusual approach,
{shown}
in Appendix A.2 in Ref.~\cite{yao}, to include the $\Delta$-contained loop Feynman Diagrams.
With this approach, the propagators of $\Delta$(1232) occurring in the loops are integrated out,
which corresponds to an expansion with respect to $1/m_\Delta$. The expansion leads to a polynomial
of $1/m_\Delta$, namely the analytic structure proportional to $\ln m_\Delta$ will never appear
in the scattering amplitude. {A direct and convenient way to compensate the contribution from $\ln m_\Delta$ terms is to adjust the values of the LECs of the tree amplitudes, since they are chiral polynomials. Actually, compared to the $O(p^4)$ fits without $\Delta$, the LECs of $O(p^4)$ in fits with $\Delta$ change a lot,
especially     {in the case of $e_{18}$.  Moreover,  }
the violation of the positivity bound is mainly caused by $e_{18}$. When the energy goes larger, bigger changes of LECs occur, possibly leading to positivity violation.} Hence, the above approach of including $\Delta$-contained loops may be practical at low energies
but invalid at high energies. However, no one knows at which energy the approach fails, as the exact full expression of the $\Delta$-contained loop amplitude is unknown.
Nevertheless,
the {positivity}  bounds can tell us something. Here, the violation of the bounds shown in Figs.~\ref{p4fitaII} indicates that the approach fails beyond
{1.2~GeV, deserving}  further calculations
of the exact $\Delta$-contained loop amplitudes.

To conclude, at $O(p^3)$ level, both the tree-level and full bounds with $\Delta$ contribution
are well satisfied, since the leading Born term of $\Delta$ gives a large and positive contribution.
At $O(p^4)$ level, the bounds are badly violated,
{which might be mainly due to the unusual way of including
the $\Delta$-contained loop contribution. The violation indicates that a further exact and full
calculation of the $\Delta$-contained loop is necessary when performing fits beyond the energy of 1.2 GeV
in the center of mass frame.}

%%%
%%%\begin{figure*}[!t]
%%%\centering
%%%\begin{minipage}{0.4\textwidth}
%%%\centering
%%%\includegraphics[clip,width=1.0\textwidth]{deltap4fitIIbtree.eps}
%%%\end{minipage}
%%%\begin{minipage}{0.4\textwidth}
%%%\centering
%%%\includegraphics[clip,width=1.0\textwidth]{deltap4fitIIbfull.eps}
%%%\end{minipage}
%%%\caption{\label{p4fitbII}Positivity bound on LECs at $O(p^4)$ level. The fit results from `Fit II(b)-$O(p^4)$'  given in Ref.~\cite{yao} are employed for plotting. Left: tree, right: tree+loop}
%%%\end{figure*}
%%%

\begin{figure*}[ht]
\centering
\begin{minipage}{0.4\textwidth}
\centering
\includegraphics[clip,width=1.0\textwidth]{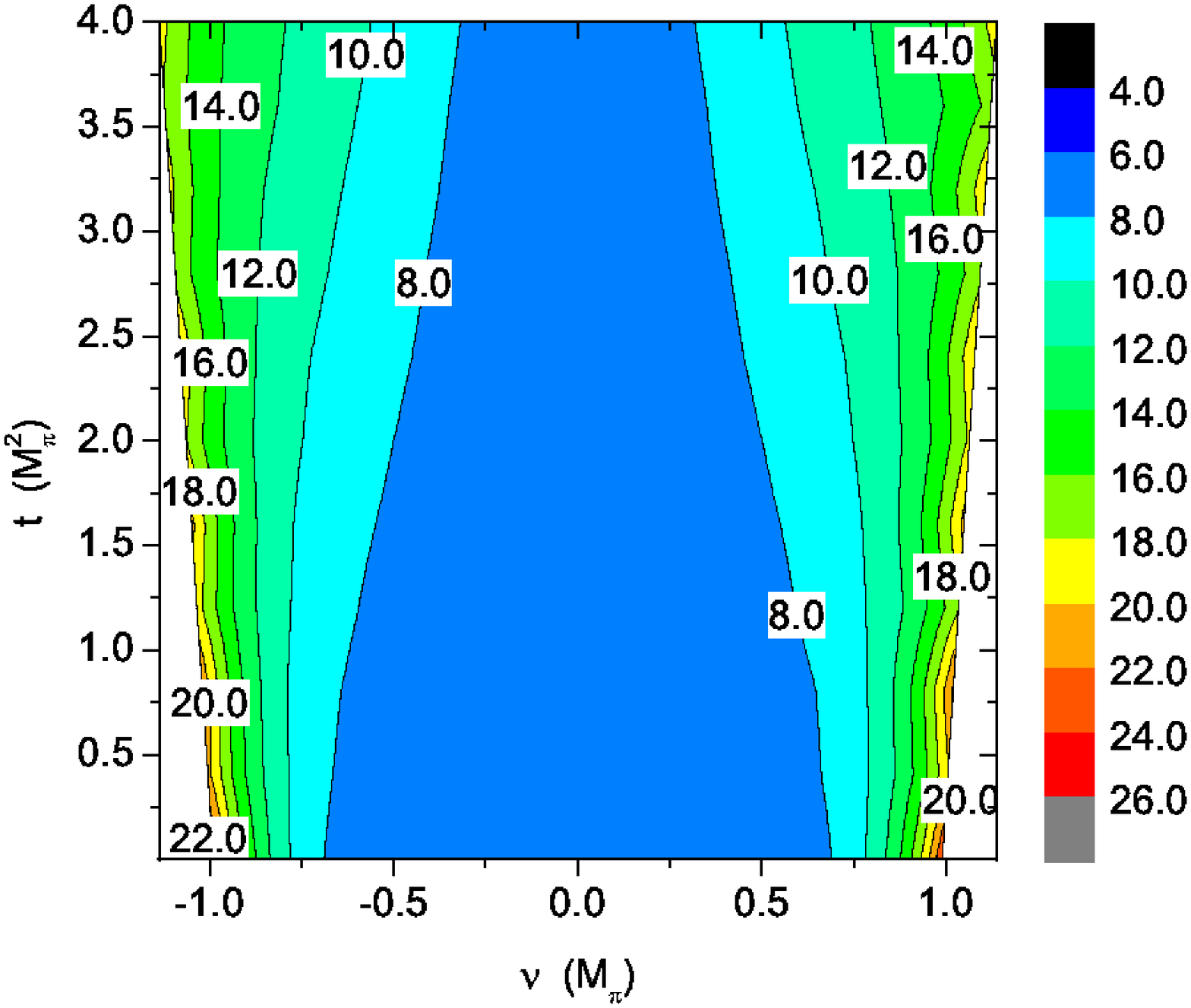}
\end{minipage}
\begin{minipage}{0.4\textwidth}
\centering
\includegraphics[clip,width=1.0\textwidth]{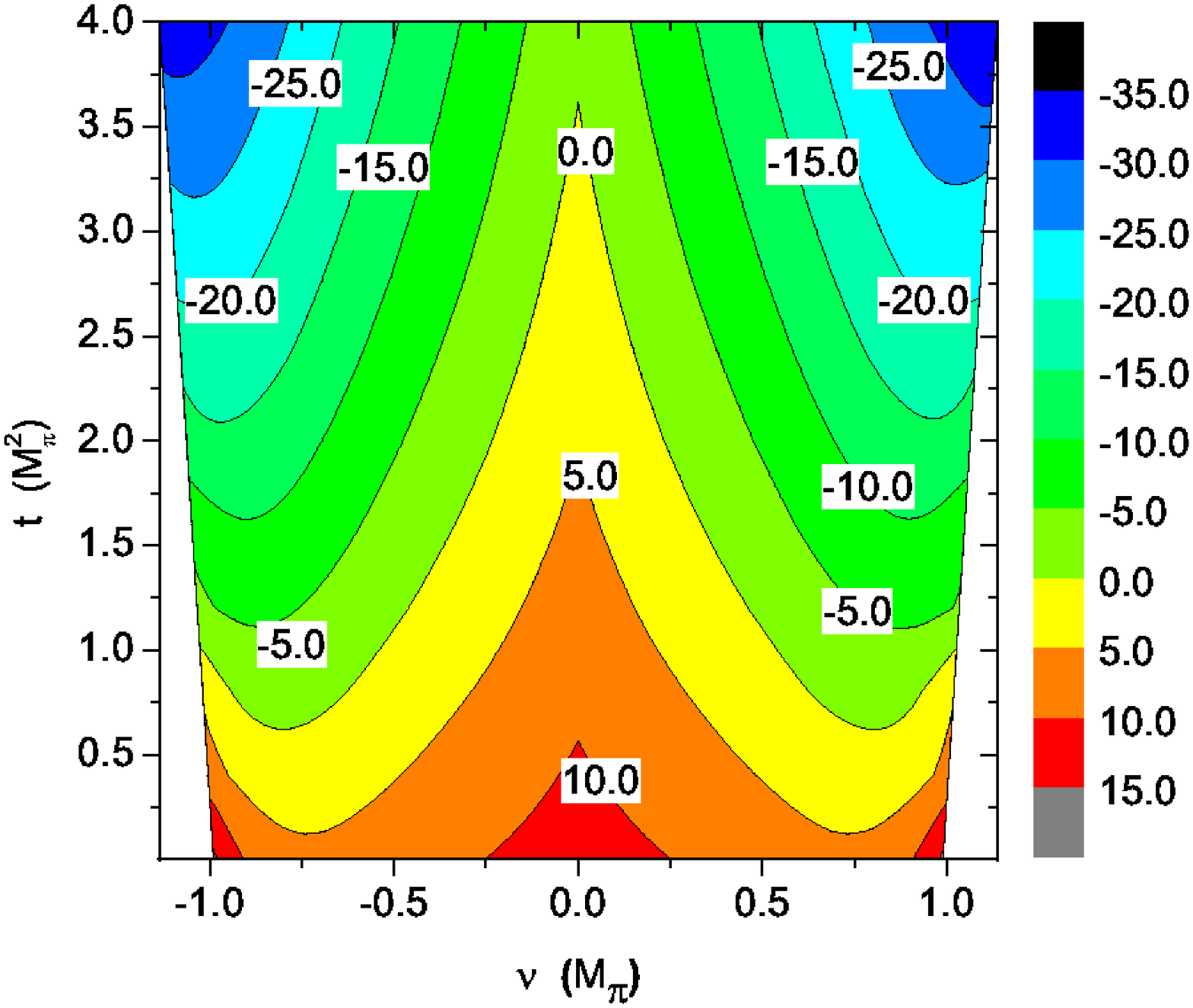}
\end{minipage}
\caption{\label{p4fitIc}\small
Positivity bound on LECs at $O(p^4)$ level. The fit results from `Fit I(c)-$O(p^4)$' and `Fit II(c)-$O(p^4)$'
given in Ref.~\cite{yao} are employed for plotting
{   $f(\alpha,\nu,t)$.   }
{\bf Left-hand side:} full bound without {explicit}
$\Delta$ contribution; {\bf right-hand side:} full bound with
{explicit} $\Delta$ contribution.
}
\end{figure*}

Finally, we would also like to discuss
{ the impact of these constraints on the values of the }
pion-nucleon sigma term,  $\sigma_{\pi N}$, analyzed in Ref.~\cite{yao}.
Therein, the lattice QCD data for $m_N$
and the pion-nucleon scattering data were employed to determine the pion-nucleon sigma term.
As a consequence of this, two different results were reported:
$\sigma_{\pi N}=52\pm7$ MeV ('Fit I(c)-$\cO(p^4)$' without
$\Delta(1232)$) and $\sigma_{\pi N}=45\pm6$ MeV ('Fit II(c)-$\cO(p^4)$' with explicit
$\Delta(1232)$ contributions).
However, though compatible, one may wonder which value is more accurate and carries less theoretical
uncertainties.
The positivity bounds derived in this work may provide an answer to  this.
%%%
%%%The two fits to obtain $\sigma_{\pi N}$ are labeled by
%%%``Fit I(c)-$O(p^4)$'' and ``Fit II(c)-$O(p^4)$'', respectively, and
%%%
The values from  ``Fit I(c)-$O(p^4)$'' and ``Fit II(c)-$O(p^4)$''   for the LECs
involved in the bounds are listed in Tabs.~\ref{tabwodel} and~\ref{tabwdel}, respectively.
The contour plots for the positivity bounds, without and  with explicit $\Delta$ contribution,
are shown in Figs.~\ref{p4fitIc}.
As we can see, the bound with $\Delta$
contribution are violated in most of the region ${\cal{R}}$, while
the one without  {explicit}   $\Delta$ contribution
are well satisfied. This may imply that the value $\sigma_{\pi N}=52\pm7$ MeV is more reasonable
than $\sigma_{\pi N}=45\pm6$ MeV.
Again we owe this to { the  lack of an exact calculation}
of the $\Delta$-contained loop.

\section{Conclusions}
\label{sec.conclusions}
Using the general S-matrix arguments, such as analyticity, crossing symmetry and unitarity,
we derived positivity constraints on the pion-nucleon scattering amplitudes
$D_\alpha(\nu,t)=\alpha A(\nu,t) +\nu B(\nu,t)$            %%%\tilde{D}+(1-\alpha)\nu \tilde{B}$
in the upper part of Mandelstam triangle, ${\cal R}$.
These constraints are further changed into positivity bounds on the chiral LECs of the pion-nucleon
Lagrangian both at $O(p^3)$ and $O(p^4)$ level.
In combination with the central values of the LECs from Refs.~\cite{yao,oller}
within EOMS-B$\chi$PT, it is found that the bounds at tree level are
always violated in some regions inside ${\cal R}$, while the full bounds (tree+loop)
are well respected both for $O(p^3)$
and $O(p^4)$ {  analyses;
loops}  are important and, in the chosen renormalization scheme (EOMS), they
{produce contributions to}
the positivity bound numerically of the same order as the tree-level diagrams.

Nonetheless, when considering the  LEC uncertainties,
the full and most stringent bounds at $O(p^3)$ level are slightly violated in some
parts of the $1\sigma$ intervals,
pointing out  the break down { of EOMS-B$\chi$PT }  for those LEC values.
However, this problem disappears  the analysis is taken up to $O(p^4)$, where
the most stringent bounds are well obeyed in the full error interval.

{We} have provided the constraints for special points where the bounds are nearly optimal
in terms of just a few $\cO(p^2)$, $\cO(p^3)$ and $\cO(p^4)$ LECs (depending on the chiral order one works at).
We hope these positivity conditions can be easily implemented
and employed to constrain future B$\chi$PT analyses.

{Finally,  the positivity bounds with an explicit $\Delta$ resonance
have been also  studied. }
The $\Delta$  Born-term provides a {positive-definite} contribution
to the bounds and hence the bounds at $O(p^3)$ level in the $\delta$-counting rule
(see Ref.~\cite{Pascalutsa}) are well satisfied. However,
at {the}  $O(p^4)$ level, the bounds are violated when
{just a part of the $\Delta$ loops is included. }
{We think that a complete one-loop calculation including $\Delta$--loops
will solve  this issue.   }

% If you have acknowledgments, this puts in the proper section head.
\begin{acknowledgments}
This work is supported in
part by National Nature Science
Foundations of China under Contract Nos. 10925522 %jieqing
%10875001%mianshang
 and
11021092,  %tuandui
{
the MICINN, Spain, under contract FPA2010-17747
and Consolider-Ingenio CPAN CSD2007-00042,
the MICINN-INFN fund AIC-D-2011-0818, the Comunidad de Madrid through
Proyecto HEPHACOS S2009/ESP-1473 and the Spanish MINECO Centro de excelencia Severo Ochoa Program
under grant SEV-2012-0249. The work of DLY was partly performed at Peking University.
}

\end{acknowledgments}

% Specify following sections are appendices. Use \appendix* if there
% only one appendix.
\appendix

\section{
\label{secspecfun}
Positive definite spectral function for    $D_\alpha^I$}

The dispersion relation for the analysis of the subthreshold amplitude is used to extract the positivity constraints.
{Hence} a positive definite spectral function in the physical region $s\geq s_{\rm th}$ is required. Starting from Eqs.~(\ref{kernelS}) and~(\ref{imageexp}), one can immediately construct a preliminary combination of the form
\bea
0 \,\leq \,  {
\rm Im}\bigg\{ \alpha_1 A^I(s,t) +\alpha_2 B^I(s,t) \bigg\} \, =\,
\vec{\alpha}^{\,\, T} \, \cdot \,  {\rm Im}\vec{\cal A}^I(s,t)\, ,
\qquad\qquad {\rm where } \quad\vec{\alpha}=\left(\ba{c} \alpha_1 \\ \alpha_2 \ea\right)\, .
\eea

Notice that in principle there is no restriction to the possible combinations we may consider,
so one may consider combinations where $\vec{\alpha}$ depends also on $s$ and $t$
or, conversely, {on}  $\nu$ and $\nu_B$.
The only necessary condition will be that they are analytical functions in the $\nu$--integration
domain in our fixed--t dispersion relation, i.e., they are real and do not contain discontinuities
for $\nu\geq \nu_{th}$ for fixed t.
Thus, for later convenience we will rather write the
general combination of $A^I$ and $B^I$ in the form
\bea
0 \,\leq \, \mbox{Im}\bigg\{  \alpha_1 A^I(s,t) +\alpha_2 \nu B^I(s,t)\bigg\} \, =\,
\vec{\alpha}^{\,\, T} \, \cdot \, \mbox{Im} \vec{\cal A}^I(s,t)\, ,
\qquad\qquad \mbox{{with} }
\vec{\alpha}=\left(\ba{c} \alpha_1 \\ \alpha_2 \nu \ea\right)\, ,
\eea
{where we introduced the $\nu$ factor in the $B^I(s,t)$ term. From  }
now on we will use the notation
\bea
D^I_\alpha(s,t) &\equiv &\vec{\alpha}^{\,\, T} \, \cdot \,  \vec{\mA}^I(s,t)\, .
\eea

For the study of the  positivity of Im$D_\alpha^I$
we will make use of the positivity of each PW, i.e.,
Im$f^I_k(s)\geq 0$ for $s\geq s_{th}$. Thus, we have that
\bea
0 \,\leq \,    \mbox{Im}D_\alpha^I(s,t)
\,&=&\,
\vec{\alpha}^{\,\, T} \, \cdot \, \sum_{\ell=0}^\infty S^{\ell}(s,t) \, \mbox{Im}\vec{\cal F}^I(s)
\nonumber\\
&&=
 \sum_{\ell=0}^\infty \vec{\alpha}^{\,\, T} \, S^{\ell}(s,t) \, \mbox{Im}\vec{\cal F}^I(s)
\nonumber\\
&&=
 \sum_{\ell=0}^\infty \bigg(\, \vec{\alpha}^{\,\, T} \vec{v}_1^{\,\, \ell}\, , \,
 \vec{\alpha}^{\,\, T}\vec{v}_2^{\,\, \ell} \, \bigg)
  \, \mbox{Im}\vec{\cal F}^I(s) \, .
\eea
For convenience, here {the} $2\times 2$ matrix $S^{\ell}(s,t)$ has been written in terms
of two dimension--2 vectors:
\bea
S^\ell(s,z_s) = \bigg( \, \vec{v}_1^{\,\, \ell}\, , \, \vec{v}_2^{\,\, \ell}\, \bigg) \, .
\eea
Hence the positivity of Im$f^I_k(s)$ ensures the positivity of Im$D_\alpha(s,t)$ whenever
\bea
\vec{\alpha}^{\,\, T} \vec{v}_1^{\,\, \ell} \geq 0\, ,  \qquad\qquad
 \vec{\alpha}^{\,\, T}\vec{v}_2^{\,\, \ell} \geq 0 \, ,
\eea
for $s\geq s_{th}$. The explicit form of these constraints is given by
\bea
\frac{4\pi}{E^2-m_N^2}\bigg( \, c_{11}(s,t) \, \alpha_1 \, +\, c_{12}(s,t)\, \nu\,\alpha_2 \bigg)
\,\geq \, 0\, ,
\nonumber\\
 -\, \frac{4\pi}{E^2-m_N^2}\bigg( \, c_{21}(s,t) \, \alpha_1 \, +\, c_{22}(s,t)\, \nu\,\alpha_2 \bigg)
 \,\geq \, 0\, ,
\eea
with
\bea
c_{11}&=& \frac{1}{2 W} \bigg[ (W+m_N)(W-W_+)(W-W_-) P'_{\ell+1}(z_s)\, +\,
 (W-m_N)(W+W_+)(W+W_-) P'_{\ell}(z_s)  \bigg] \, ,
 \nonumber\\
c_{12}&=& (E-m_N)  P'_{\ell+1}(z_s)\, - \,
 (E+m_N) P'_{\ell}(z_s) \, ,
 \nonumber\\
c_{21}&=& \frac{1}{2 W} \bigg[ (W+m_N)(W-W_+)(W-W_-) P'_{\ell}(z_s)\, +\,
 (W-m_N)(W+W_+)(W+W_-) P'_{\ell+1}(z_s)  \bigg] \, ,
 \nonumber\\
c_{22}&=& (E-m_N)  P'_{\ell}(z_s)\, - \,
 (E+m_N) P'_{\ell+1}(z_s) \, ,
 \nonumber\\
\eea
{and the kinematical variables,
\bear
&& z_s(t,s) = 1+\Frac{t}{2\vec{q}^{\, \, 2}}\, , \qquad \qquad
|\vec{q}| = \sqrt{\Frac{\lambda(s,M_\pi^2,m_N^2)}{4 s}}\, ,
\nn\\
&& W=\sqrt{s}\, , \qquad \qquad
E=\sqrt{\vec{q}^{\,\, 2} +m_N^2 } = \Frac{W^2 +m_N^2-M_\pi^2}{2 W}    \, ,
\nn\\
&& W_\pm^2 = (m_N\pm M_\pi)^2\, , \qquad \qquad
s_{th}=s_+\, ,
\eear     }
{with $\vec{q}$ being the three-momentum of the pion in the center-of-mass
rest-frame.   }

Since $E\geq m_N$ when $s\geq s_{th}$ we can simplify the inequalities in the form
\bea
 c_{11}  \, \alpha_1 \, +\, c_{12} \,\nu\, \alpha_2
\,\geq \, 0\, ,
\nonumber\\
 c_{21}  \, \alpha_1 \, +\, c_{22} \, \nu\, \alpha_2
 \,\leq \, 0\, .
\eea

The coefficients $c_{mn}$ are combinations of the first derivative of the Legendre polynomials
and in general the sign may change from one partial wave $\ell$ to another $\ell'$, or from an
energy $(s,t)$ to another.   However, when $z_s(s,t)\geq 1$, i.e., when $t\geq 0$ for $s\geq s_{th}$,
one has that $P'_k(z_s)\geq 0$ and then
\bea
c_{11}\geq 0\, , \qquad c_{21}\geq 0\, ,
\eea
for any $s\geq s_{th}$ and $t\geq 0$  (as $W\geq m_N\geq 0$ and $W\geq W_\pm>0$).
Thus, the inequalities { get simplified } into the form
\bea
&&\alpha_1 \, \geq \, -\, \frac{ c_{12}}{c_{11}} \, \nu\, \alpha_2\ ,
\qquad\alpha_1 \, \leq \, -\, \frac{ c_{22}}{ c_{21}  } \, \nu\, \alpha_2
\ .\nonumber\\
\eea
One can further simplify this expression by means of the relations
$(E\pm m_N)= (W\pm W_+)(W\pm W_-)/(2W)$. {We can then  }
write the inequalities in the form
\bea
\alpha_1 \, \geq \, -\, \alpha_2\, \frac{\nu}{W+m_N}\, \left[\frac{ 1 - g^\ell_+(s,t)
}{  1+\frac{(W-m_N)}{(W+m_N)}   g^\ell_+(s,t) } \right]
\, ,
\nonumber\\
\alpha_1 \, \leq \,  \alpha_2\, \frac{\nu}{W- m_N}\,  \left[  \frac{  1 - g^\ell_-(s,t)
}{  1+\frac{(W+m_N)}{(W-m_N)}   g^\ell_-(s,t) }  \right]   \, ,
\eea
with
\bea
g^\ell_\pm(s,t)  &=&  \frac{(W\pm W_+)(W\pm W_-)}{(W\mp W_+)(W\mp W_-)}
\, \frac{P'_{\ell}(z_s)}{P'_{\ell+1}(z_s)}\, .
\eea
Notice that these functions depend not only on the energy $(s,t)$ but also on the PW index $\ell$.
Hence, we will have to obtain the region obtained by the overlap of all the PW constraints.
The analysis of the Legendre {polynomials}  tells us that for $z_s\geq 1$,
\bea\label{legendre}
\frac{P'_\ell(z_s)}{P'_{\ell+1}(z_s)} \, < \,\frac{P'_{\ell+1}(z_s)}{P'_{\ell+2}(z_s)}
\, < \, \lim_{\ell\to\infty}
\frac{P'_\ell(z_s)}{P'_{\ell+1}(z_s)} \,=\, \frac{1}{z_s \, +\, \sqrt{z_s^2-1}} \, .
\eea
Thus, we can define the upper-bound  functions  for $t\geq 0$ and $s\geq s_{\rm th}$
(which implies $z_s\geq 1$),
\bea
\bar{g}_\pm(s,t)  &=&   \frac{(W\pm W_+)(W\pm W_-)}{(W\mp W_+)(W\mp W_-)}
\,\left[  \frac{1}{z_s \, +\, \sqrt{z_s^2-1} }\right]\,\,\,
\geq \,\,\, g^\ell_\pm(s,t) \, .
\eea
Hence, the intersection of all the PW's $\ell$ is given by the
{most stringent}  constraints for $t\geq 0$ and
$s\geq s_{th}$,   {given by the limit functions $\bar{g}_\pm(s,t)$:    }
\bea
\alpha_1 \, \geq \, -\, \alpha_2\, \frac{\nu}{W+m_N}\, \left[\frac{ 1 - \bar{g}_+(s,t)
}{  1+\frac{(W-m_N)}{(W+m_N)}   \bar{g}_+(s,t) } \right]
\, ,
\nonumber\\
\alpha_1 \, \leq \,  \alpha_2\, \frac{\nu}{W- m_N}\,  \left[  \frac{  1 - \bar{g}_-(s,t)
}{  1+\frac{(W+m_N)}{(W-m_N)}   \bar{g}_-(s,t) }  \right]   \, .
\nonumber\\
\eea
These {two} constraints {have} (at least) an allowed region in the quadrant $\alpha_{1,2}\geq 0$,
bounded by the two straight lines provided by these inequalities.

Now we proceed to the analysis of the bounds for the variable $s\geq s_{th}$.
One can see that the most stringent constraints come from  the range when
\bea
 -\,  \frac{\nu}{W+m_N}\, \left[\frac{ 1 - \bar{g}_+(s,t)
}{  1+\frac{(W-m_N)}{(W+m_N)}   \bar{g}_+(s,t) } \right]\,
\eea
is maximum and when
\bea
\frac{\nu}{W- m_N}\,  \left[  \frac{  1 - \bar{g}_-(s,t)
}{  1+\frac{(W+m_N)}{(W-m_N)}   \bar{g}_-(s,t) }  \right]   \,
\eea
is minimum.
For fixed $t$ one can check that the respective maximum and minimum are always found
for $s=s_{th}$. Thus, the most restricted region among all $s\geq s_{th}$ for fixed--$t\geq 0$
is given  by
\bea
\alpha_1 \, \geq \,  \alpha_2\,
\alpha_{\rm min}(t) \, , \qquad\qquad
\alpha_1 \, \leq \,  \alpha_2\, \alpha_{\rm max}(t)\, ,
\eea
with
\bea
\alpha_{\rm min}(t) &=& \lim_{s\to s_{th}}
( -1)\times   \frac{\nu}{W+m_N}\, \left[\frac{ 1 - \bar{g}_+(s,t)
}{  1+\frac{(W-m_N)}{(W+m_N)}   \bar{g}_+(s,t) } \right]\,
= \, \frac{(4 m_N^2 -t )(4 m_N M_\pi+t) }{4 m_N ( 4 m_N^2 M_\pi + 2 m_N t +M_\pi t) }
\nonumber\\&&
\qquad\qquad \,=\, 1 \, -\, \frac{t}{4 m_NM_\pi}
{\rm {
\, +\,
\Frac{ t(t-4 M_\pi^2)}{8 M_\pi^2 m_N^2}
\,\, \, +\,\, \,
{\cal O}\bigg(\frac{p^3}{m_N^3}\bigg) \, ,
}}
\nonumber\\
\alpha_{\rm max}(t)&=& \lim_{s\to s_{th}}
\frac{\nu}{W- m_N}\,  \left[  \frac{  1 - \bar{g}_-(s,t)
}{  1+\frac{(W+m_N)}{(W-m_N)}   \bar{g}_-(s,t) }  \right]
\,
=\, 1 \, + \, \frac{t}{4 m_NM_\pi} \, ,
\eea
{
with $M_\pi=\cO(p)$ and $t=\cO(p^2)$~\cite{yao}. For $0\leq t\leq 4 M_\pi^2$
one has  $\alpha_{\rm min}(t)\leq 1- t/(4 m_N M_\pi)$.
}

Taking into account that the Mandelstam triangle, free of analytical cut-singularities,
is given by {   $s\leq (m_N+M_\pi)^2$, $u\leq (m_N+M_\pi)^2$   }
and $t\leq 4 M_\pi^2$,
in combination with our positivity assumption $t\geq 0$,
we find that only combinations with $\alpha_1\geq 0$ and $\alpha_2\geq 0$ are allowed
so, up to a global  {irrelevant} positive number $\alpha_2$
the constraints finally become   (after relabeling
{  $\alpha_1$ as $\alpha\,\alpha_2$)  }
\bea
\alpha_{\rm min}(t) \,\leq \, \alpha \,\leq \, \alpha_{\rm max}(t)\, .
\eea
Notice that we have optimized the bound for $\alpha$ for every $\ell$ and $s\geq s_{th}$.
Thus, finally, this condition ensures the positivity of the spectral function combination
for $s\geq s_{th}$ and $t\ge 0$,
\bea
\mbox{Im}D_\alpha^I(s,t)\,\geq 0\, .
\eea

The combination can then be written in the form
\bea
D^I_\alpha &=& \alpha A^I \,+\, \nu  B^I \,\,\,=\,\,\, \alpha D^I \,+\, (1-\alpha)\, \nu  B^I\, ,
\eea
{where $D_1^I(\nu,t)$ is equal to the usual $D^I(\nu,t)$.  }

% Create the reference section using BibTeX:
%\bibliography{basename of .bib file}

\end{document}